\documentclass[useAMS,usenatbib]{mn2e}
\usepackage{amsmath}
\usepackage{amssymb}
\usepackage{graphicx}

\newcommand{\dv}{\ensuremath{\Delta \vec{v}}}

\title[Constraining the Stellar Mass Function in the Galactic Center Via Mass Loss from Stellar Collisions]{Constraining the Stellar Mass Function in the Galactic Center Via Mass Loss from Stellar Collisions}
\author[Douglas Rubin and Abraham Loeb]{Douglas Rubin$^{1}$\thanks{E-mail:
dsrubin@physics.harvard.edu (DSR); aloeb@cfa.harvard.edu (AL)} and Abraham Loeb$^{2}$\\
$^{1}$Department of Physics, Harvard University, Cambridge, MA 02138, USA\\
$^{2}$Department of Astronomy, Harvard University, Cambridge, MA 02138, USA}
\begin{document}

\date{Accepted farther in the future. Received in the future; in original form }

\pagerange{\pageref{firstpage}--\pageref{lastpage}} \pubyear{2010}

\maketitle

\label{firstpage}

\begin{abstract}
 The dense concentration of stars and high velocity dispersions in the Galactic centre imply that stellar collisions frequently occur.  Stellar collisions could therefore result in significant mass loss rates.  We calculate the amount of stellar mass lost due to indirect and direct stellar collisions and find its dependence on the present-day mass function of stars.  We find that the total mass loss rate in the Galactic centre due to stellar collisions is sensitive to the present-day mass function adopted.  We use the observed x-ray luminosity in the Galactic centre to preclude any present-day mass functions that result in mass loss rates $> 10^{-5} \mathrm{M_{\odot} yr^{-1}}$ in the vicinity of $\sim 1''$.  For present-day mass functions of the form, $dN/dM \propto M^{-\alpha}$, we constrain the present-day mass function to have a minimum stellar mass $\lesssim 7\mathrm{M_{\odot}}$ and a power law slope $\gtrsim 1.25$.  We also use this result to constrain the initial mass function in the Galactic centre by considering different star formation scenarios.

\end{abstract}


\section{Introduction}

The dense stellar core at the Galactic centre has a radius of $\sim 0.15 - 0.4 \mathrm{pc}$, a stellar density $> 10^6\mathrm{M_{\odot} pc^{-3}}$ \citep{genzel:1996, eckart:1993, genzel:2003, schodel:2007} high velocity dispersions ($\ge 100\mathrm{km} \, \mathrm{s^{-1}}$), and Sgr A*, the central supermassive black hole with a mass $\approx 4 \times 10^{6}M_{\odot}$ \citep{eckart:2002, schodel:2002, schodel:2003, ghez:2003, ghez:2008}.  Due to the extreme number densities and velocities, stellar collisions are believed to play an important role in shaping the stellar structure around the Galactic centre, and in disrupting the evolution of its stars.  \citet{genzel:1996} found a paucity of the brightest giants in the galactic center, and proposed that collisions with main sequence (MS) stars could be the culprit.  This hypothesis was found to be plausible by \citet{alexander:1999}.  Other investigations of collisions between giants and MS, white dwarf and neutron stars \citep{bailey:1999} and collisions between giants and binary MS and neutron stars \citep{davies:1998} could not account for the dearth of observed giants.  The contradictory results were resolved by \citet{dale:2009}, who concluded that the lack of the faintest giants (but not the brightest giants) could be explained by collisions between giants and stellar mass black holes.  Significant mass loss in the giants' envelopes after a collision would prevent the giants from becoming bright enough to be observed.

The above studies concentrated on collisions involving particular stellar species with particular stellar masses.  To examine the cumulative effect of collisions amongst an entire ensemble of a stellar species with a spectrum of masses, one must specify the present-day stellar mass function (PDMF) for that species.  The PDMF gives the current number of stars per unit stellar mass up to a normalization constant.  Given a certain star formation history, the PDMF can be used to determine the initial mass function of stars (IMF), the mass function with which the stars were born.  There is currently no consensus as to whether the IMF in the Galactic centre deviates from the canonical IMF \citep{bastian:2010}.

First described by Salpeter more than 50 years ago \citep{salpeter:1955}, the canonical IMF is an empirical function which has been found to be universal \citep{kroupa:2001}, with the Galactic centre as perhaps the sole exception.  \citet{maness:2007} found that models with a top-heavy IMF were most consistent with observations of the central parsec of the Galaxy.  \citet{paumard:2006}, and subsequently \citet{bartko:2010} found observational evidence for a flat IMF for the young OB-stars in the Galactic centre.  On the other hand, \citet{lockmann:2010} concluded that models of constant star formation with a canonical IMF could explain observations of the Galactic centre.

In this work we use calculated mass loss rates due to stellar collisions as a method to constrain the PDMF for main sequence stars in the Galactic centre.  We construct a simple model to estimate the actual mass loss rate in the Galactic centre based on observed x-ray emission.  PDMFs that predict mass loss rates from stellar collisions greater than the observed rate are precluded.  This method allows us to place conservative constraints on the PDMF, because we do not include the contribution to the mass loss rate from stellar winds from massive evolved stars \citep{baganoff:2003}.  Specifically, this method allows us to place a lower limit on the power-law slope and an upper limit on the minimum stellar mass of the PDMF in the Galactic centre (see \S~\ref{sec:constrain}).  Inclusion of the mass loss rate from stellar winds (or other sources) could further constrain the PDMF of the Galactic centre.

We present novel, analytical models to calculate the amount of stellar mass lost due to stellar collisions between main sequence stars in \S~\ref{sec:condition} through \S~\ref{sec:mass_loss_direct}.  In \S~\ref{sec:coll_rates} we develop the formalism for calculating collision rates in the Galactic center.  We utilize our calculations of the mass loss per collision, and the collision rate as a function of Galactic radius to find the radial profile of the mass loss rate in \S~\ref{sec:mass_loss_rates}.  Since the amount of mass lost is dependent on the masses of the colliding stars, the mass loss rate in the Galactic centre is sensitive to the underlying PDMF.  By comparing our calculations to mass loss rates obtained from the x-ray luminosity measured by \textit{Chandra}, in \S~\ref{sec:constrain} we constrain the PDMF of the Galactic centre.  We derive analytic solutions of the PDMF as a function of an adopted IMF for different star formation scenarios, which allows us to place constraints on the IMF in \S 6.  In \S~7, we estimate the contribution to the mass loss rate from collisions involving red giant (RG) stars.

\section{Condition for Mass loss}
\label{sec:condition}

Throughout this paper we refer to the star that loses material as the perturbed star, and the star that causes material to be lost as the perturber star.  Quantities with the subscript or superscript ``pd'' or ``pr'' refer to the perturbed star and perturber star respectively.  \footnote{Note that for any particular collision, it is arbitrary which star we consider the perturber star, and which star the perturbed star.  Both stars will lose mass due to the presence of the other, so in order to calculate the total mass loss, we interchange the labels (pd$\leftrightarrow$pr), and repeat the calculation.}   We work in units where mass is measured in the mass of the perturbed star, $M_{pd}$, distance in the radius of the perturbed star, $r_{pd}$, velocity in the escape velocity of the perturbed star, $v_{esc}^{pd}$ ($= \sqrt{2GM_{pd}/r_{pd}}$), and time in $r_{pd}/v_{esc}^{pd}$.  We denote normalization by these quantities (or the appropriate combination of these quantities) with a tilde:
 \begin{eqnarray}
\tilde M &\equiv &M/M_{pd} \nonumber \\
\tilde r &\equiv &r/r_{pd}  \nonumber  \\
\tilde v &\equiv &v/v_{esc}^{pd}  \nonumber  \\
\tilde t & \equiv &t/\frac{r_{pd}}{v_{esc}^{pd}}.  \nonumber  \\
\end{eqnarray}
We refer to collisions in which $b > r_{pd}+r_{pr}$ as ``indirect'' collisions, and collisions in which $b\leq r_{pd}+r_{pr}$ as ``direct'' collisions.  The impact parameter, b,  is the distance of closest approach measured from the centers of both stars.

We consider the condition for mass loss at a position, $\tilde r$, within the perturbed star to be that the kick velocity due to the encounter at $\tilde r$ exceeds the escape velocity of the perturber star at $\tilde r$, $\Delta \tilde v (\tilde r) \geq \tilde{v}_{esc}(\tilde r)$.  The escape velocity as a function of position within the perturbed star can be found from the initial kinetic and potential energies of a test particle at position $\tilde r$,
\begin{eqnarray}
\label{eqn:v_esc}
\tilde{v}_{esc} (\tilde{r}) &=& \sqrt{-\int_{\tilde r}^{\infty}\frac{ \tilde M_{int}(\tilde r^{\prime})}{\tilde r^{\prime 2}}d\tilde r^{\prime}} \nonumber \\
&=&  \sqrt{\frac{\tilde M_{int}(\tilde r)}{\tilde r} + 4\pi \int_{\tilde r}^{1} \tilde \rho (\tilde r^{\prime}) \tilde r^{\prime} d\tilde r^{\prime}}
\end{eqnarray}
where $\tilde M_{int}$ is the mass interior at position $\tilde {r}^{\prime}$ and $\tilde \rho$ is the density profile of the star.

\subsection{Mass loss due to Indirect Collisions}
\label{sec:kick_vel}

To calculate the mass lost due to an indirect collision, we first calculate the kick velocity given to the perturbed star as a function of position within the star.  We work under the impulse approximation \citep{spitzer:1958}, valid under the condition that the encounter time is much shorter than the characteristic crossing time of a constituent of the perturbed system.

Given a mass distribution for the perturbed system, $\rho_{pd}$ and a potential for the perturber system, $\Phi$, the kick velocity after an encounter under the impulse approximation is given by~\citet{binney:2008}:
\begin{align}
\dv (\vec{r}) & = -\int_{-\infty}^{\infty} \left [ \vec{\nabla} \Phi (\vec{r}, t) \right. \nonumber \\
&-\frac{1}{M_{pd}}\int \rho_{pd}(\vec{r^{\prime}}, t) \vec{\nabla} \Phi(\vec{r^{\prime}}, t)\,d^3r^{\prime}\left. \vphantom{\vec{\nabla} \Phi (\vec{r}, t)} \right ] \,dt.
\label{eqn:dv_impulse}
\end{align}
 Equation~(\ref{eqn:dv_impulse}) can be simplified by expanding the gradient of the potential in a Taylor series, resulting in
 \begin{equation}
 \dv\,(\vec{r}) = \frac{2GM_{pr}}{b^2v_{rel}}\begin{pmatrix} -x\\y\\0 \end{pmatrix} + O\left (\overline{r^2} \right).
 \label{eqn:dv_tide}
 \end{equation}
The expansion is valid under the ``distant tide'' approximation which is satisfied when $r_{pd} \ll b$.  The parameter $v_{rel}$ is the relative speed between the stars ($v_{rel} \equiv |\vec{v}_{pd}-\vec{v}_{pr}|$).  We are interested in the magnitude of equation~(\ref{eqn:dv_tide}), which when normalized to the units that we have adopted for this paper is
 \begin{equation}
\label{eq:dv_units}
\Delta \tilde v ( \tilde x, \tilde y ) \cong \gamma \sqrt {\tilde x^2+\tilde y^2 },
\end{equation}
where

\begin{equation}
\label{eq:def_of_gamma}
\gamma \equiv \frac{\tilde M_{pr}}{\tilde b^2 \tilde{v}_{rel}}.
\end{equation}

To solve for the mass lost per encounter as a function of $\gamma$, we consider a star within a cubic array, where the star contains $\sim 3 \times 10^{6}$ cubic elements.  As a function of $\gamma$ we compare the kick velocity in each element to the escape velocity for that element, and consider the mass within the element to be lost to the star if the velocities satisfy the condition given \S~\ref{sec:condition}.  We note that by $\sim 10^5$ elements, the results converge to within about $2\%$, and we are therefore confident that $\sim 3 \times 10^{6}$ provides adequate resolution.

To calculate the amount of mass in each element, the density profile for the perturbed star must be specified.  As with several previous studies on mass loss due to stellar collisions \citep{benz:1987, benz:1992, lai:1993, rauch:1999} we utilize polytropic stellar profiles.  Polytropic profiles are easy to calculate, and yield reliable results for stars of certain masses.  Polytropic profiles of polytropic index $n=1.5$ describe the density structure of fully convective stars, and therefore very well describe MS stars with $M_{\star} \lesssim 0.3 \mathrm M_{\odot}$ (nearly fully convective) and MS stars with $M_{\star} \gtrsim 10\mathrm M_{\odot}$ (convective cores).  MS stars with $M_{\star} \gtrsim 1\mathrm M_{\odot}$ have radiative envelopes, and are therefore well described by $n=3$.  For $n$ for stars with masses of $0.3 - 1\mathrm M_{\odot}$ and $5 - 10\mathrm M_{\odot}$, we linearly interpolate between $n=1.5$ and 3.  We discuss the uncertainties introduced by this approach in \S~\ref{sec:mass_loss_rates}.  Note that this approach is biased towards zero-age main sequence stars, since as stars evolve, they are less adequately described by polytropic profiles.

\begin{figure}
\centerline{\includegraphics[clip, width=3.2in, angle=-90]{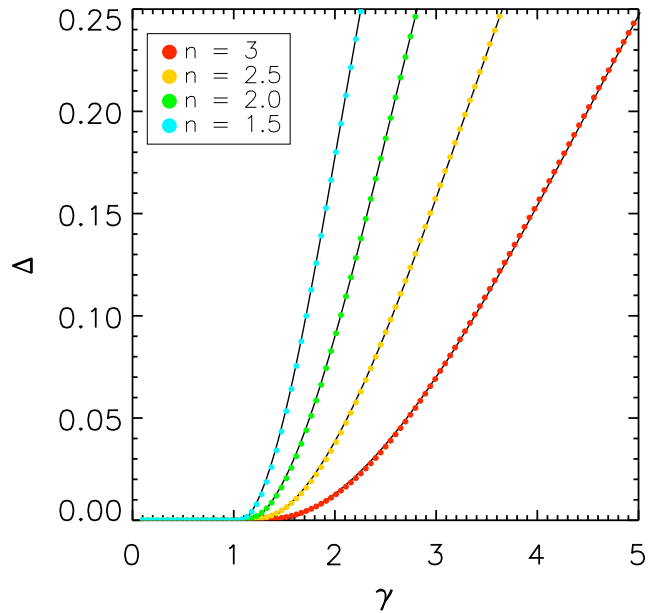}}
\caption{ \label{fig:mass_loss}The fraction of mass lost per collision as a function of $\gamma$ for several polytropic indeces.  The lines are third order polynomial fits, whose coefficients are given in Table~\ref{tab:coeffs}.}
\end{figure}

We plot the fraction of mass lost from the perturbed star per event, $\Delta$, as a function of $\gamma$ in Fig.~\ref{fig:mass_loss} for several polytropic indeces.  The lines are third order polynomial fits to our results, in the range of $0.98 \leq \gamma \leq 5$.  We list the coefficients of the polynomial fits in Table~\ref{tab:coeffs}.  For each density profile, no mass is lost up until $\gamma$ of about 0.98, and thereafter the mass loss increases monotonically.  The increasing trend is due to the fact that larger perturber masses and smaller impact parameters result in an increased potential felt by the perturbed star.  Smaller velocities also cause more mass to be lost, as this increases the ``interaction time'' between the perturber and perturbed stars.

    \begin{table}
    \centering
    \caption{\label{tab:coeffs}Coefficients of polynomial fits for $\Delta(\gamma)$ with varying polytropic indeces}
    \begin{tabular}{ccccc}
      \hline
      \hline
       $n$ & $a_{0}$ & $a_{1}$ & $a_{2}$ & $a_{3}$ \\
      \hline
1.5 & 0.395 & -0.865 & 0.559 & -0.091 \\
2.0 & 0.210 & -0.424 & 0.246 & -0.032 \\
2.5 & 0.105 & -0.197 & 0.102 & -0.101 \\
3.0 & 0.051 & -0.088 & 0.040 & -0.003 \\
      \hline
      \hline
    \end{tabular}
  \end{table}

\begin{figure*}
\centerline{\includegraphics[clip, width=5.1in]{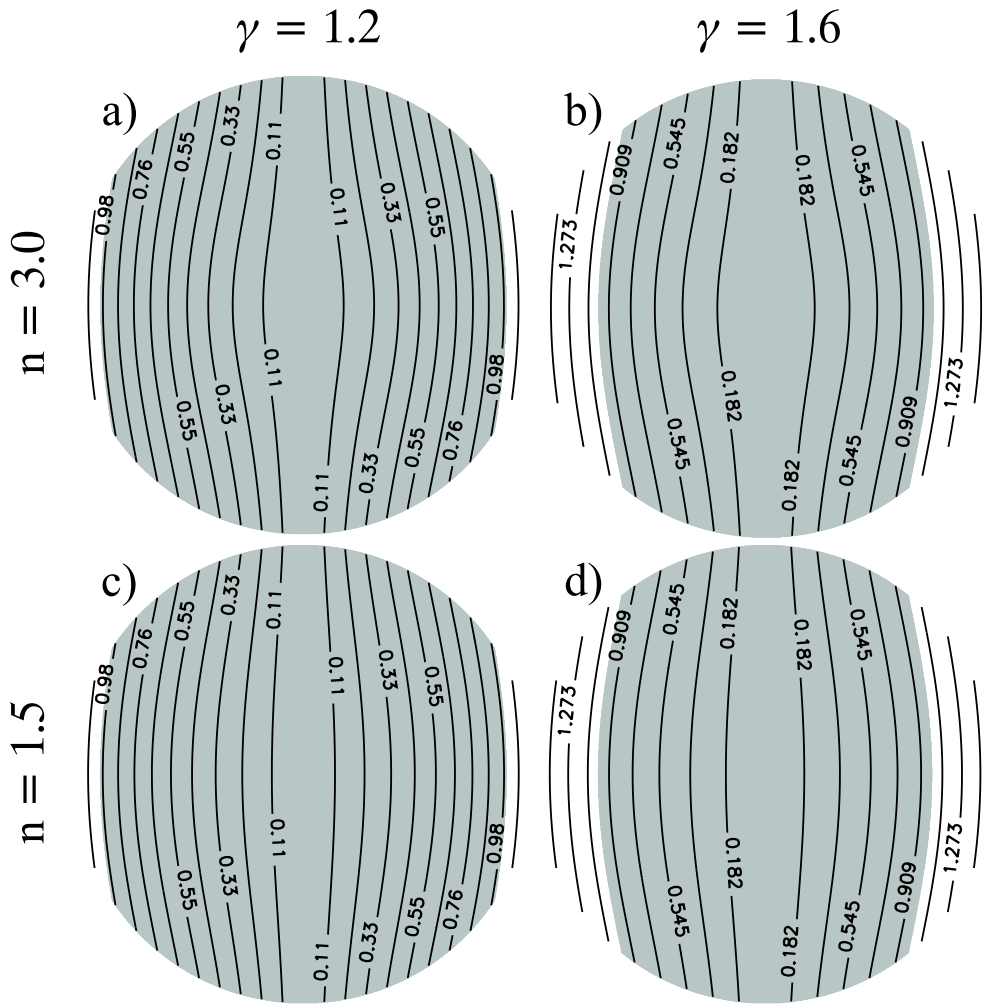}}
\caption{ \label{fig:slices_mass_lost}  Slices through the perturbed star along the plane parallel to the perturber star's trajectory.  The first column (panels a) and c)) correspond to encounters with $\gamma = 1.2$, and the second column (panels b) and d)) correspond to $\gamma = 1.6$. The first row (panels a) and b)) have $n=3$, and the second row (panels c) and d)) have $n=1.5$. The contours are the kick velocity within the star due to the encounter normalized to the escape velocity (as a function of $\tilde r$). The outline of the grey region underneath has $\Delta \tilde v/\tilde v_{esc}^{pd}(\tilde r) = 1$, so that the grey region represents the location of where mass is still left after the event.}
\end{figure*}

The location of the mass loss within the perturbed star for fixed $\gamma$ depends upon the polytropic index, since the escape velocity within the star is dependent upon the density profile, as indicated by equation.~(\ref{eqn:v_esc}).  In Fig.~\ref{fig:slices_mass_lost}, we illustrate where mass will be lost in the perturbed star by plotting contours of the kick velocity ($\Delta \tilde v (\tilde r)$) due to the encounter normalized to $\tilde v_{esc}^{pd} (\tilde r)$ for $n=1.5$ and $n=3$ (top and bottom row respectively).  We show two different cases: a slightly perturbing encounter with $\gamma=1.2$ in the first column, and a severely perturbing encounter with $\gamma=1.6$ in the second column.  The grey region underneath shows where mass is still left after the encounter, since $\Delta \tilde v/\tilde v_{esc}^{pd} (\tilde r)$ within this region is $< 1$.  The $\gamma=1.6$ encounter results in bigger kick velocities, and so we see that the mass loss penetrates farther into the star.  We note that the shape and magnitude of the contours for both polytropic indeces at fixed $\gamma$ converge at large radii.  This is due to the fact that regardless of the polytropic index used, $\tilde v_{esc}^{pd}$ converges to the same value at large radii when the second term in equation~(\ref{eqn:v_esc}) becomes negligible.  Even though the location of where mass is lost is similar for different polytropic stars at the same value of $\gamma$, the amount of the mass lost is substantially different (as shown in Fig.~\ref{fig:mass_loss}), due to the different density profiles.

\subsection{Validity of approach for indirect collisions}
\label{sec:validity}

The impulse approximation is valid provided that the time over which the encounter takes place, $t_{enc}$, is much shorter than the time it takes to cross the perturbed system, $t_{cross}$.  To estimate when our calculations break down, we approximate $t_{enc}$ as $b/v_{rel}$, and $t_{cross}$ as $t_s$, the time it takes for a sound wave to cross an object that is in hydrostatic equilibrium:
\begin{equation}
t_{cross}\sim t_s \sim \frac{1}{\sqrt{G\bar{\rho}_{pd}}}\sim \frac{1}{\sqrt{GM_{pd}/r_{pd}^3}}.
\end{equation}
These approximations lead to the condition that
\begin{equation}
\label{eq:impulse_cond}
\tilde{v}_{rel}^{-1} \tilde b \ll 1.
\end{equation}

\begin{figure}
\centerline{\includegraphics[clip, width=3.5in]{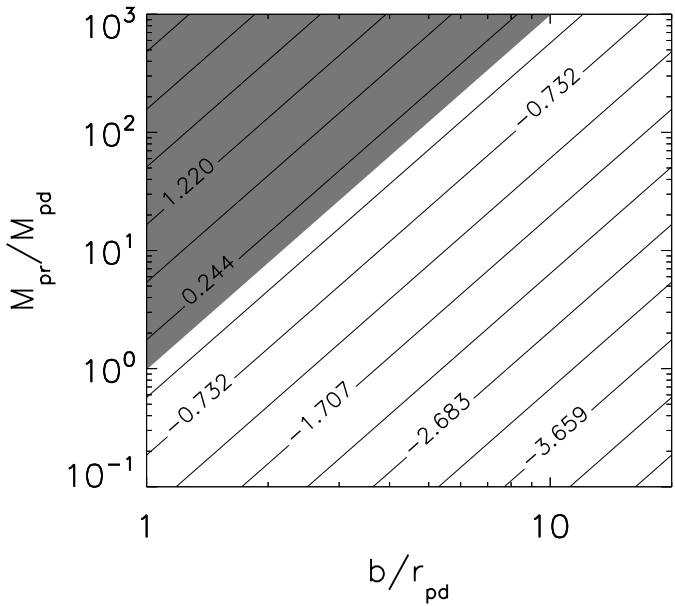}}
\caption{ \label{fig:impulse_valid}Contours of log($\gamma_{valid}$) in the $M_{pr}/M_{pd}$ - $b/r_{pd}$ parameter space, where $\gamma_{valid}$ is defined in \S~\ref{sec:condition}.  The shaded grey area indicates where the impulse approximation predicts non-zero mass loss}.
\end{figure}

\citet{aguilar:1985} find that for a large range of collisions, the impulse approximations remains remarkably valid, even when $t_{enc}$ is almost as long as $t_{cross}$.  We therefore assume that the impulse approximation holds until the left hand side of equation (\ref{eq:impulse_cond}) is $\sim 1$.   Our calculation of $\Delta$ as a function of $\gamma$ should therefore be valid for $\gamma \lesssim \gamma_{valid}$, where $\gamma_{valid} \equiv \tilde M_{pr}/\tilde b^3$.  We plot contours of log$(\gamma_{valid})$ in the $M_{pr}/M_{pd}$ - $b/r_{pd}$ parameter space in Fig.~\ref{fig:impulse_valid}, where both the x and y axes span ranges relevant to our calculations.  The shaded grey area in the figure is the region of the parameter space where the impulse approximation predicts non-zero mass loss due to the encounter.  The figure shows that $\gamma_{valid}$ is smaller for low $M_{pr}$ to $M_{pd}$ ratios at high impact parameters.  In fact, most of the right side of the parameter space has $\gamma_{valid} $ less than 0.98 (where below this value, the impulse approximation predicts no mass lost).

In our calculations, when, for any particular set of $M_{pr}/M_{pd}$ and $b/r_{pd}$, $\gamma >\gamma_{valid}$, we adopt $\Delta(\gamma>\gamma_{valid}) = \Delta(\gamma=\gamma_{valid})$.  This approach represents a lower limit on the amount of mass loss that we calculate, since mass loss should increase with increasing $\gamma$.  We find, however, that if we set $\Delta(\gamma > \gamma_{valid}) =1$ (which represents the absolute upper limit in the amount of mass lost) the change in our final results is negligible at small Galactic radii.  At large radii, where the mass loss from indirect collisions dominates (see \S~\ref{sec:mass_loss_rates}), the results change by at most a fact of $\sim2$.

Equation~(\ref{eqn:dv_tide}) was derived under the assumption that the impact parameter is much bigger than both $r_{pd}$ and $r_{pr}$.  Since $\Delta v$ scales as $b^{-2}$, the equation predicts that most mass loss occurs for small impact parameters.  However, given the assumption that was used to derive the equation, the regime of small impact parameters is precisely where equation.~(\ref{eqn:dv_tide}) breaks down.  Numerical simulations \citep{aguilar:1985, gnedin:1999} show that for a variety of perturber mass distributions, the energy input into the perturbed system is well described by equation~(\ref{eqn:dv_tide}) for $b \gtrsim 5 r_h$, where $r_h$ is the half mass radius of the perturber system.  For an $n=3$ polytropic star, $5r_h=1.4r_{\star}$.  Since for indirect collisions, $b/r_{pd} = 1+r_{pr}/r_{pd}+d/r_{pd}$ (where d is the distance between the surface of both stars), there is only a small region in our calculations, $0 \le d/r_{pd} \lesssim (0.4 - r_{pr}/r_{pd})$, for which equation~(\ref{eqn:dv_tide}) may give unreliable results.

\subsection {Mass loss due to direct Collisions}
\label{sec:mass_loss_direct}

A number of papers over the past few decades have addressed the outcomes of stellar collisions where the two stars come so close to each other that not only gravitational, but also hydrodynamic forces must be accounted for.  Early studies used one or two dimensional low resolution hydrodynamic simulations (e.g. Mathis 1967; DeYoung 1968).  Modern studies typically utilize smooth particle hydrodynamics with various stellar models, mass-radius relations, and varying degrees of particle resolution \citep{benz:1987, benz:1992, lai:1993, rauch:1999}.  A detailed review of the literature can be found in this area can be found in \citet{freitag:2005}.

We approach the problem of direct collisions in a highly simplified, analytic manner without hydrodynamic considerations, and find that for determining the amount of mass lost, our method compares well to the complex hydrodynamic simulations.  As a first order model, we approximate the encounter as two colliding disks, by projecting the mass of both stars on a plane perpendicular to the trajectory of the perturber star.  The problem of calculating mass loss then becomes easier to handle, as it is two dimensional.  We also assume that mass loss can only occur in the geometrical area of intersection of the two stars.

We find the kick velocity as a function of position in the area of intersection by conserving momenta, and by assuming that all of the momentum in the perturber star in each area element was transfered to the corresponding area element in the perturbed star.  Working in the frame of the perturbed star, and with a polar coordinate system at its center (so that $r=\sqrt{x^2+y^2}$), we find

\begin{equation}
\Delta \tilde v (\tilde r) = \frac{\tilde \Sigma_{pr}(\tilde r)\tilde v_{rel}}{\tilde \Sigma_{pd}(\tilde r)}.
\end{equation}
The parameters $\Sigma_{pr}$ and $\Sigma_{pd}$ represent the surface density of the perturber and perturbed stars respectively ($\Sigma \equiv \int \rho dz$).

  \begin{figure}
\centerline{\includegraphics[clip, width=3.4in]{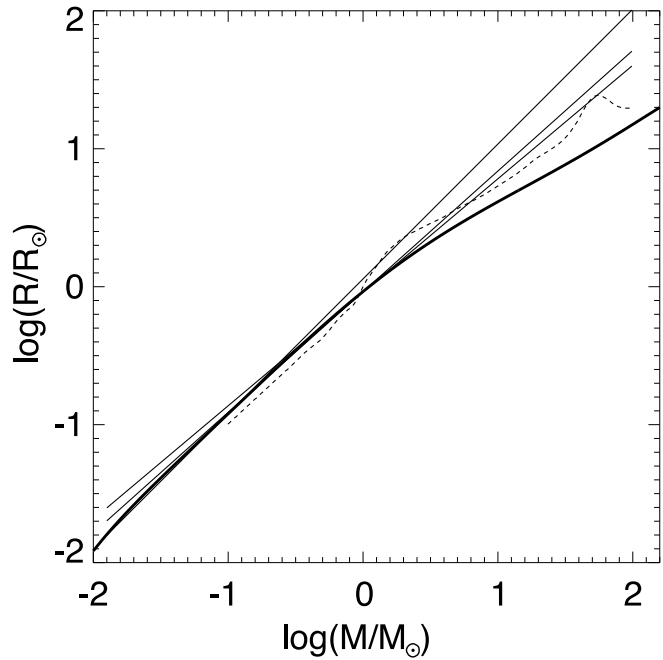}}
\caption{ \label{fig:m_r_relation}Mass-radius relations used in studies of calculating mass loss from stars due to stellar collisions.  The thin lines are power-law relations of power-law index 1.0, 0.8, and 0.85 used by  \citet{rauch:1999}, \citet{lai:1993} and \citet{benz:1992} respectively.  The dotted line is the relation used by \citet{freitag:2005}, and the thick line is the relation used in this work.}
\end{figure}

To find the region of intersection, we need to know the impact parameter, and the radii of both stars.   To obtain the stellar radii as a function of mass, we use the mass-radius relation calculated by \citet{kippenhahn:1994} for a MS star with $Z=Z_{\odot}$, $X_H = 0.685$, and $X_{He} = 0.294$ from a stellar evolution model, where $X$ represents the mass fraction.  We fit a polynomial to their Fig. 22.2, and extrapolate on the high and low mass ends so that we have a mass-radius relation that spans from about 0.01 to 150M$_{\odot}$.  We compare our mass-radius relation to those used in other studies of direc stellar collisions in Fig.~\ref{fig:m_r_relation}.  \citet{rauch:1999}, \citet{lai:1993} and \citet{benz:1992} all adopted power laws with power law indices of 1.0, 0.8, 0.85 respectively (thin lines).  \citet{freitag:2005} (dotted line) use main sequence stellar evolution codes to obtain a mass-radius relation for masses $> 0.4 \mathrm M_{\odot}$, and a polytropic mass-radius relation of $n=1.5$ for masses $< 0.4\mathrm M_{\odot}$.

\begin{figure*}
\centerline{\includegraphics[clip, width=7.5in]{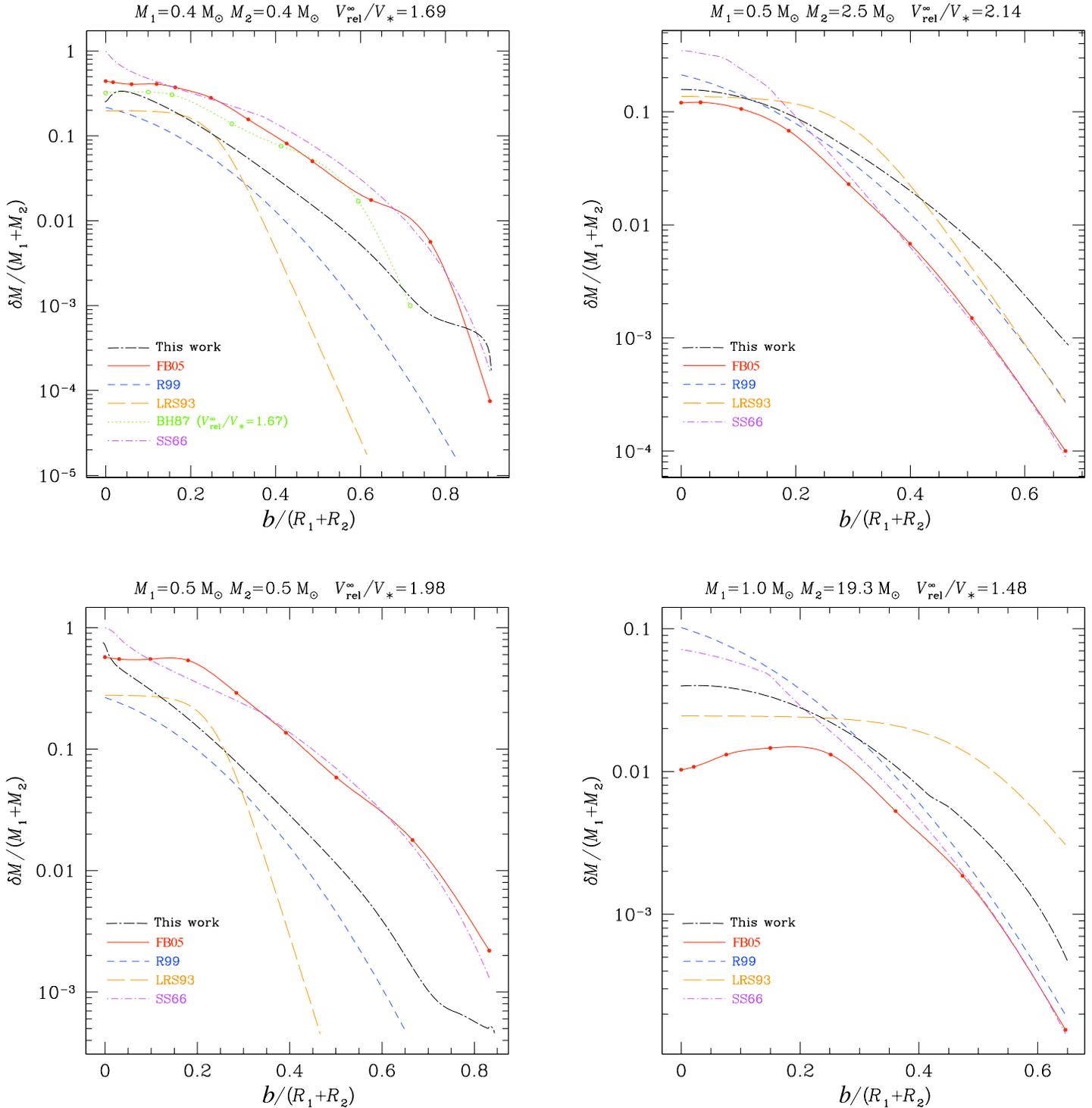}}
\caption{ \label{fig:comp_1}The calculated fractional amount of mass lost as a function of impact parameter from several works.  Our results are the black dashed-dotted lines.  The acronyms FB05, R99, LRS93, BH87, and SS66 refer to \citet{freitag:2005}, \citet{rauch:1999}, \citet{lai:1993}, \citet{benz:1987} and \citet{spitzer:1966} respectively.  The figures are adopted from \citet{freitag:2005}.}
\end{figure*}

\begin{figure*}
\centerline{\includegraphics[clip, width=7.5in]{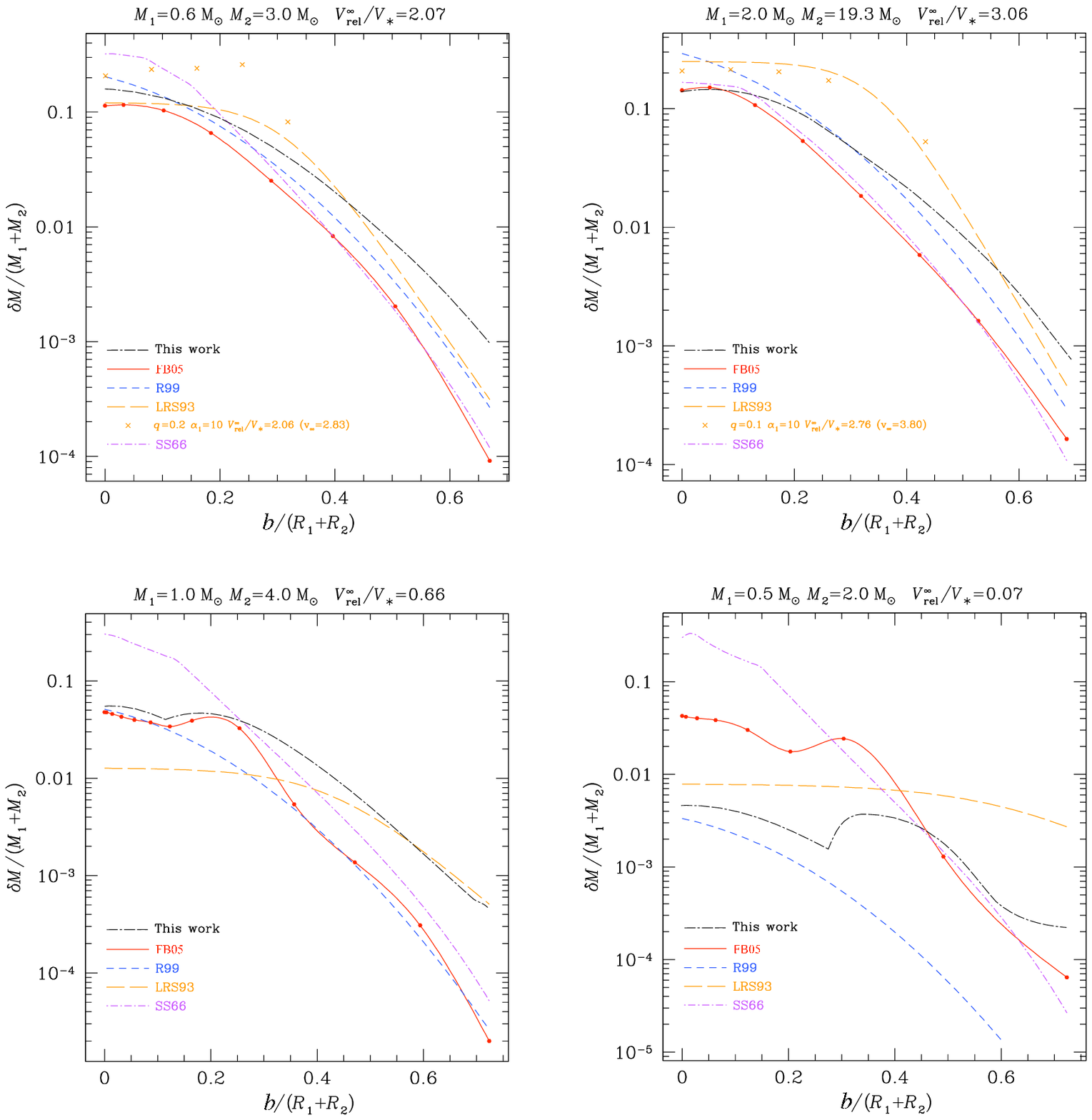}}
\caption{ \label{fig:comp_2}The calculated fractional amount of mass lost as a function of impact parameter from several works.  Our results are the black dashed-dotted lines.  The figures are adopted from \citet{freitag:2005}.}
\end{figure*}

Our simple model for calculating mass loss due to direct stellar collisions compares surprisingly well to full blown smooth particle hydrodynamic simulations.  We borrow plots of the fractional amount of mass lost as a function of impact parameter for specific relative velocities and stellar masses from \citet{freitag:2005} (Figs.~\ref{fig:comp_1} and~\ref{fig:comp_2}).  They show their own work, the best calculations of mass loss due to stellar collisions to date.  For comparison, and to show how the calculations have evolved over the years, the results from older studies are also shown.  Our own results are plotted (dashed-dotted black lines) over these previous studies.  We make sure to show results spanning a wide range of stellar masses and relative velocities.  Note that these plots show the fractional amount of mass lost from both stars normalized to the initial masses of both stars, and that the impact parameter is normalized to the sum of both stellar radii.  Our results show the same qualitative trends seen in the \citet{freitag:2005} curves, even replicating several ``bumps" seen in their curves (see the two bottom panels of Fig.~\ref{fig:comp_2}).  As compared to the \citet{freitag:2005} results, for any specific set stellar masses, relative velocity and impact parameter, our calculations sometimes over or under-predict the amount of mass lost by of a factor of a few to at most a factor of 10.
We discuss the error introduced into our main calculations by this discrepancy at the end of \S\ref{sec:constrain}.

\section{Stellar collision rates in the Galactic Center}
\label{sec:coll_rates}

To calculate mass loss rates in the Galactic center, we will need to find the collision rates as a function of the perturber and perturbed star masses, impact parameter, and relative velocity.  Additionally, the collision rate will be a function of distance from the Galactic center, since the stellar densities and relative velocities vary with this distance.  In this section, we first present the Galactic density profile that we use, and we then derive the differential collision rate as a function of these parameters.

We adopt the stellar density profile of \citet{schodel:2007}, one of the best measurements of the density profile within the Galactic centre to date.  Using stellar counts from high resolution images of the galactic center, they find that the density profile is well approximated by a broken power law.  Moreover, they use measured velocity dispersions to constrain the amount of enclosed stellar mass as a function of galactic radius, $r_{gal}$.  Using their density profile, and velocity dispersion measurements, they find that

\begin{align}
\label{eq:rho_enc}
& \bar{\rho} (r_{gal}) = \nonumber \\
&   \left \{
 \begin{array}{lcl}
2.8 \pm 1.3 \times 10^{6}M_{\odot}pc^{-3}  \left ( \frac{r_{gal}}{0.22pc} \right )^{-1.2}  &\mbox{for }  r_{gal}\leq 0.22~pc\\
2.8 \pm 1.3 \times 10^{6}M_{\odot}pc^{-3}  \left ( \frac{r_{gal}}{0.22pc} \right )^{-1.75} &\mbox{for } r_{gal}> 0.22~pc
\end{array} \right. .
\end{align}
Their average density can be converted into a local density, $\rho(r_{gal})$, by considering the definition of $\bar{\rho}$,
\begin{equation}
\bar{\rho}(r_{gal}) \equiv \frac{\int_0^{r_{gal}} 4 \pi r_{gal}^{\prime 2} \rho(r_{gal}^{\prime})dr_{gal}^{\prime}}{4/3\pi r_{gal}^3},
\end{equation}
from which we derive:
\begin{equation}
\label{eq:conv_rho_enc_to_rho}
\rho(r_{gal}) = \bar{\rho}(r_{gal})+\frac{r_{gal}}{3}\frac{d\bar{\rho}(r_{gal})}{dr_{gal}}.
\end{equation}

\begin{figure}
\centerline{\includegraphics[clip, width=3.4in]{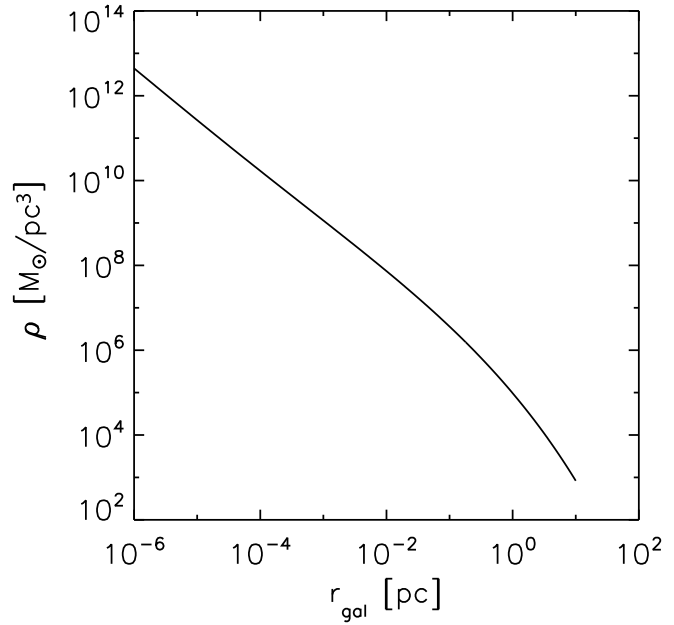}}
\caption{ \label{fig:rho_v_r} The stellar density profile that we adopt, based on the average density profile of \citet{schodel:2007}.}
\end{figure}

We use equations~(\ref{eq:rho_enc}) and~(\ref{eq:conv_rho_enc_to_rho}), to find $\rho(r)$, and plot the result in Fig.~\ref{fig:rho_v_r}.  We ``smoothed'' the unphysical discontinuity in $\rho$ arising from the kink of the broken power law fit by fitting a polynomial to equation~(\ref{eq:rho_enc}).

The differential collision rate, $d\Gamma$, between two species, ``1'' and ``2'' at impact parameter b characterized by distribution functions $f_{1}$ and $f_{2}$, and moving with relative velocity $|\vec{v}_1-\vec{v}_2|$ in a spherically symmetric system is

\begin{align}
\label{eq:d_gamma_1}
 d\Gamma &= f_{1}(r_{gal}, \vec{v}_{1})\,d^3 v_{1}f_{2}(r_{gal},\vec{v}_{2})\,d^3v_{2} \nonumber \\
 & \times |\vec{v}_{1} - \vec{v}_{2}| 2\pi b \,db \,4 \pi r_{gal}^2 dr_{gal}.
 \end{align}
For simplicity, we adopt Maxwellian distributions,
\begin{equation}
 f_{1,2} (r_{gal}, \vec{v}_{1,2})= \frac{n_{1,2}(r_{gal})}{(2\pi \sigma^2)^{3/2}}e^{-v_{1,2}^2/2\sigma^2},
\end{equation}
where we find the velocity dispersion, $\sigma$, from the Jean's equations.  Assuming an isotropic velocity dispersion, a spherical distribution of stars, and a power-law density profile with power-law slope $\beta$, $\rho \propto r_{gal}^{-\beta}$, the Jean's equations lead to $\sigma^2 = \frac{GM_{\mathrm{SMBH}}}{r_{gal}(1+\beta)}$, where $M_{SMBH} = 4\times10^6\mathrm M_{\odot}$.  From Fig.~\ref{fig:rho_v_r}, it is evident that $\beta(r_{gal})$, but for simplicity, we adopt an averaged value of $\beta$, $\beta =1.3$.  Note that we have also assumed that the enclosed mass at position $r_{gal}$ is dominated by the SMBH.  This assumptions is valid out till $\sim1pc$, which is also the point where our impulse approximation starts to break down.

A change of variables allows one to integrate out 3 of the velocity dimensions, and to write the expression in terms of $v_{rel}$ (see Binney \& Tremaine 2008).  We can also take into account the fact that both species have a distribution of masses by introducing, $\xi_{1,2}$, the PDMF, which gives the number density of stars per mass bin ($\xi \equiv dn/dM$).  We adopt a power law PDMF,
\begin{equation}
 \xi \propto M^{-\alpha},
 \end{equation}
 that runs from some minimum mass, $M_{min}$ to a maximum mass $M_{max}$.  Since most initial mass functions are parameterized with a power law, the present-day mass function might be modified from a power law due to the effects of collisions and stellar evolution.  Although the actual PDMF might have deviations from a power law, adopting a power law provides us with a quick and simple way to parameterize the PDMF.  Taking all of this into account, and assuming that the relative velocities are isotropic, we arrive at the final non-dimensionalized expression for the differential collision rate:
\begin{align}
\label{eq:dgamma_final}
d\tilde{\Gamma} &= 4\pi^{3/2}\tilde{\sigma}^{-3} e^{-\tilde{v}_{rel}^2/4\tilde{\sigma}^2} \tilde{v}_{rel}^3 \tilde{K}^2 \nonumber \\ & \times \tilde{M}_1^{-\alpha} \tilde{M}_2^{-\alpha} \tilde{r}_{gal}^2\tilde{b}\,d\tilde{b}\,d\tilde{r}_{gal}\,d\tilde{v}_{rel}\, d\tilde{M}_1\,d\tilde{M}_2.
\end{align}
The tildes denote normalization by the proper combination of $M_2$, $r_2$, and $v_{esc}^{2}$. The parameter K is the normalization constant for $\xi$, which can be solved for by using the density profile of Fig.~\ref{fig:rho_v_r} and the following expression:
\begin{align}
\rho & =\int_{M_{min}}^{M_{max}} \frac{dn}{dM}M dM \nonumber \\
&=K(r_{gal})\int_{M_{min}}^{M_{max}} M^{1-\alpha} dM \nonumber \\
&=K(r_{gal})\frac{M_{max}^{2-\alpha}-M_{min}^{2-\alpha}}{2-\alpha}.
\end{align}
Since the expression for K, which controls the total number of stars, has no time dependence, our expression for the PDMF assumes a constant star formation rate in the Galactic center.

Our calculations involve the computation of multidimensional integrals over a two dimensional parameter space (see \S \ref{sec:mass_loss_rates}).  Therefore, for the ease of calculation, we ignore the enhancement of the collision rate due to the effects of gravitational focusing.  This results in a conservative estimate of the collision rate.  As two projectiles collide with each other, their mutual gravitational attraction pulls them together, resulting in an enhancement of the cross section:
\begin{equation}
\label{eq:grav_focusin}
 S \rightarrow S\left (1+\frac{2G(M_1+M_2)}{bv_{rel}^2}\right).
 \end{equation}
We discuss the uncertainties in our final results due to ignoring gravitational focusing at the end of \S \ref{sec:constrain}.

\begin{figure*}
\centerline{\includegraphics[clip, width=7.in]{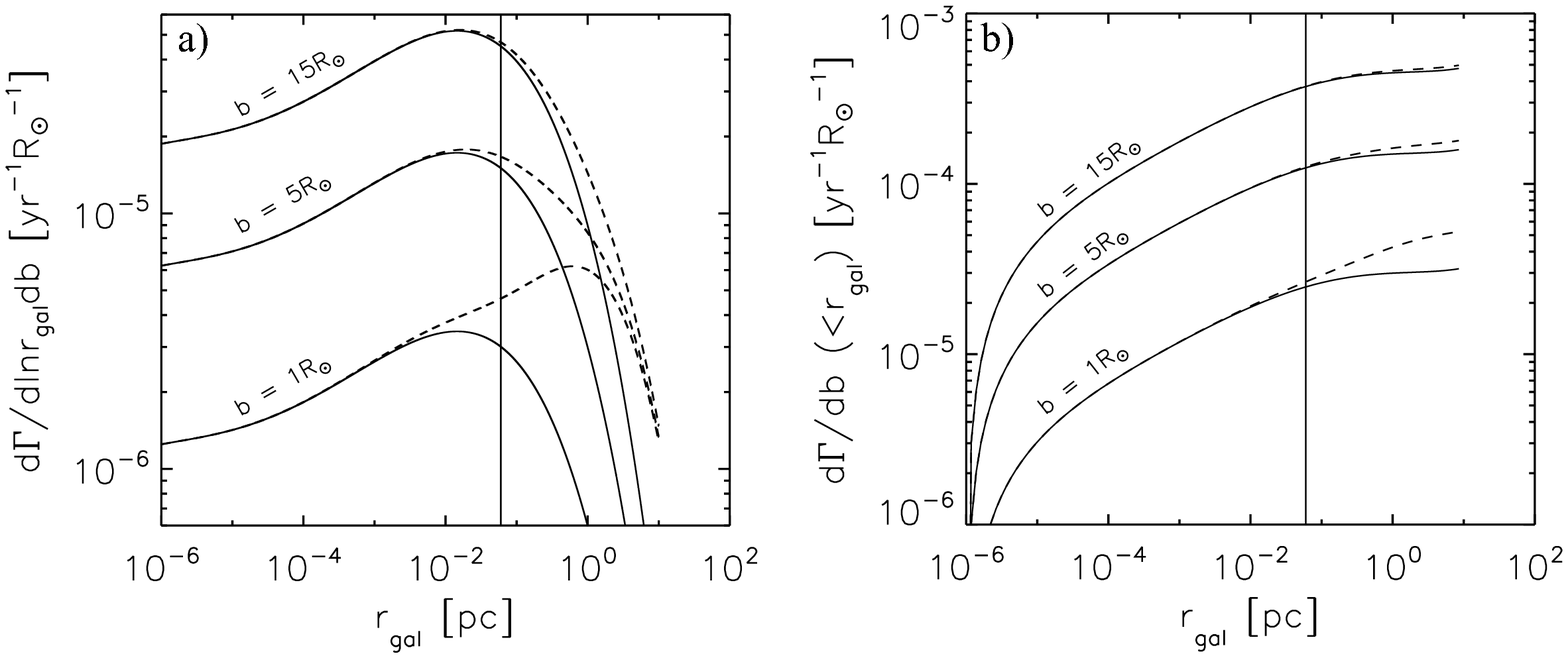}}
\caption{ \label{fig:diff_coll_rate} a) The differential collision rate per logarithmic Galactic radius per impact parameter as a function of Galactic radius for several different impact parameters.  The solid (dashed) lines were calculated ignoring (including) gravitational focusing.  The curves were made by made by integrating equation (\ref{eq:dgamma_final}) (with and without the gravitational focusing term) assuming Salpeter values. b) The cumulative differential collision rate (integrated over $r_{gal}$) per impact parameter with and without gravitational focusing for the same impact parameter values.  a and b) The vertical line in each panel is placed at $r_{gal} = 0.06$pc, the upper bound in our integration across $r_{gal}$ as performed in \S \ref{sec:constrain}.}
\end{figure*}

To illustrate the frequency of collisions in the Galactic center, we integrate equation.~(\ref{eq:dgamma_final}) over $v_{rel}$, $M_1$ and $M_2$ assuming a Salpeter-like mass function ($\alpha =2.35$, $M_{min} =0.1M_{\odot}$ and $M_{max} =125M_{\odot}$) to obtain $d\Gamma/(dlnr_{gal}db)$ as a function of $r_{gal}$ (Fig.~\ref{fig:diff_coll_rate}a)\footnote{This figure, and subsequent figures in this paper with $r_{gal}$ as the independent variable start from $r_{gal} = 10^{-6}$pc.  This value of $r_{gal}$ corresponds to the tidal radius for a $1\mathrm{M_{\odot}}$ star associated with a $4 \times 10^6$M$_{\odot}$ SMBH.  Although stars of different masses will have slightly different tidal radii, the main conclusions of our paper are based off of distances in $r_{gal}$ of order 0.1pc (see \S \ref{sec:constrain}), well above the tidal radius for any particular star.}.  We plot $d\Gamma/(dlnr_{gal}db)$ for several different impact parameter values.  We calculate $d\Gamma/(dlnr_{gal}db)$ with and without the effect of gravitational focusing (solid and dashed lines respectively).  The latter is obtained by multiplying equation (\ref{eq:dgamma_final}) by the gravitational focusing enhancement term before the integration.  As expected, gravitational focusing is negligible at small Galactic radii since typical stellar encounters involve high relative velocities.  As the typical relative velocities decrease with increasing Galactic radius, the enhancement to the collision rate from gravitational focusing becomes important.  The figure also shows that gravitational focusing becomes less important with increasing impact parameter since the gravitational attraction between the stars is weaker.  Fig.~\ref{fig:diff_coll_rate}b shows the cumulative differential collision rate (integrated over $r_{gal}$) per impact parameter as a function of $r_{gal}$.  Again, we plot the results with and without gravitational focusing and for the same impact parameters.

\section{Mass loss rates in the Galaxy}
\label{sec:mass_loss_rates}

To calculate the mass loss rate from stars due to collisions within the Galactic center, we multiply equation~(\ref{eq:dgamma_final}) by the fraction of mass lost per collision, $\Delta(\gamma)$, and compute the multi-dimensional integral.  We calculate the total mass loss rate from both the perturbed and perturber stars by simply interchanging the ``pr'' and ``pd'' labels and re-performing the calculation.

 We first compute the differential mass loss rate for indirect collisions.  The mass loss per collision is given by:
\begin{equation}
\label{eq:mass_loss}
\Delta (\gamma) =  \left \{ \begin{array}{ccl}
0 & for & \gamma<  0.98 \\
polynomial & for & 0.98 \le \gamma \le \gamma_{valid} \\
\Delta(\gamma_{valid}) & for & \gamma> \gamma_{valid} \end{array}\right. .
\end{equation}
The coefficients for the polynomial depend on the polytopic index of the perturbed star (and thus on its mass) and are taken from Table~\ref{tab:coeffs}.  We multiply equation~(\ref{eq:mass_loss}) and equation~(\ref{eq:dgamma_final}) and simplify the integration.  In principle, b should go to $\infty$, but we cut off the integral at $\tilde b_{max}=20$ as we find that the results converge well before this point.  The velocity integral is also cut off at $\tilde v_{max}$ due to the fact that $\Delta (\gamma)$ becomes zero below $\gamma = 0.98$.  This cut-off corresponds to $\tilde v_{max} = \frac{(\tilde M_{pr})_{max}}{0.98\tilde b_{min}^2}$.  We may safely throw away the exponential as  $\tilde v_{max}^2 \ll \tilde \sigma^2(\tilde r_{gal})$ for the range of $\tilde r_{gal}$ that we consider.  Thus, the integral that we evaluate is:
\begin{align}
\label{eq:mass_loss_indir}
\left(\frac{d\tilde{\dot{M}}}{dln\tilde{r}_{gal}}\right)_{pd} & \cong 4 \pi^{3/2} \tilde{\sigma}^{-3} \tilde{r}_{gal}^3 \tilde{K}^2\int_{\tilde{M}_{min}}^{\tilde{M}_{max}} \int_{\tilde{M}_{min}}^{\tilde{M}_{max}}\int_{0}^{\tilde{v}_{max}} \nonumber \\
& \times \int_{1+\tilde{r}_{pr}}^{\tilde{b}_{max}} \tilde{b}\tilde{v}_{rel}^3\Delta_{pd}(\gamma) \tilde{M}_{pr}^{-\alpha}\tilde{M}_{pd}^{-\alpha}d\tilde{b} d\tilde{M}_{pr}  \nonumber \\
& \times d\tilde{M}_{pd} d\tilde{v}_{rel}.
\end{align}

For direct collisions, $\Delta(\tilde{b}, \tilde{M}_{pr}, \tilde{M}_{pd}, \tilde{v}_{rel})$ is calculated given the prescription in Sec.\ref{sec:mass_loss_direct}.  To evaluate the multidimensional integral, we make the approximation of evaluating $\Delta_{pd}$ at $\tilde{v}_{rel}=2\tilde{\sigma}$.  The factor of $\Delta_{pd}$ thus comes out of the $\tilde{v}_{rel}$ integral, so that the $\tilde{v}_{rel}$ integral can be performed analytically:
\begin{align}
\label{eq:mass_loss_dir}
\left(\frac{d\tilde{\dot{M}}}{dln\tilde{r}_{gal}}\right)_{pd} & \cong 32 \pi^{3/2} \tilde{\sigma} \tilde{r}_{gal}^3 \tilde{K}^2\int_{\tilde{M}_{min}}^{\tilde{M}_{max}} \int_{\tilde{M}_{min}}^{\tilde{M}_{max}} \int_{0}^{1+\tilde{r}_{pr}} \nonumber \\
& \times  \tilde{b} \Delta_{pd}(\tilde{b}, \tilde{M}_{pr}, \tilde{M}_{pd}, \tilde{v}_{rel}=2\tilde{\sigma}(\tilde{r}_{gal}))\nonumber \\
& \times \tilde{M}_{pr}^{-\alpha}\tilde{M}_{pd}^{-\alpha}d\tilde{b} d\tilde{M}_{pr}d\tilde{M}_{pd}.
\end{align}
We evaluate the remaining integrals numerically.

\begin{figure*}
\centerline{\includegraphics[clip, width=7.1in]{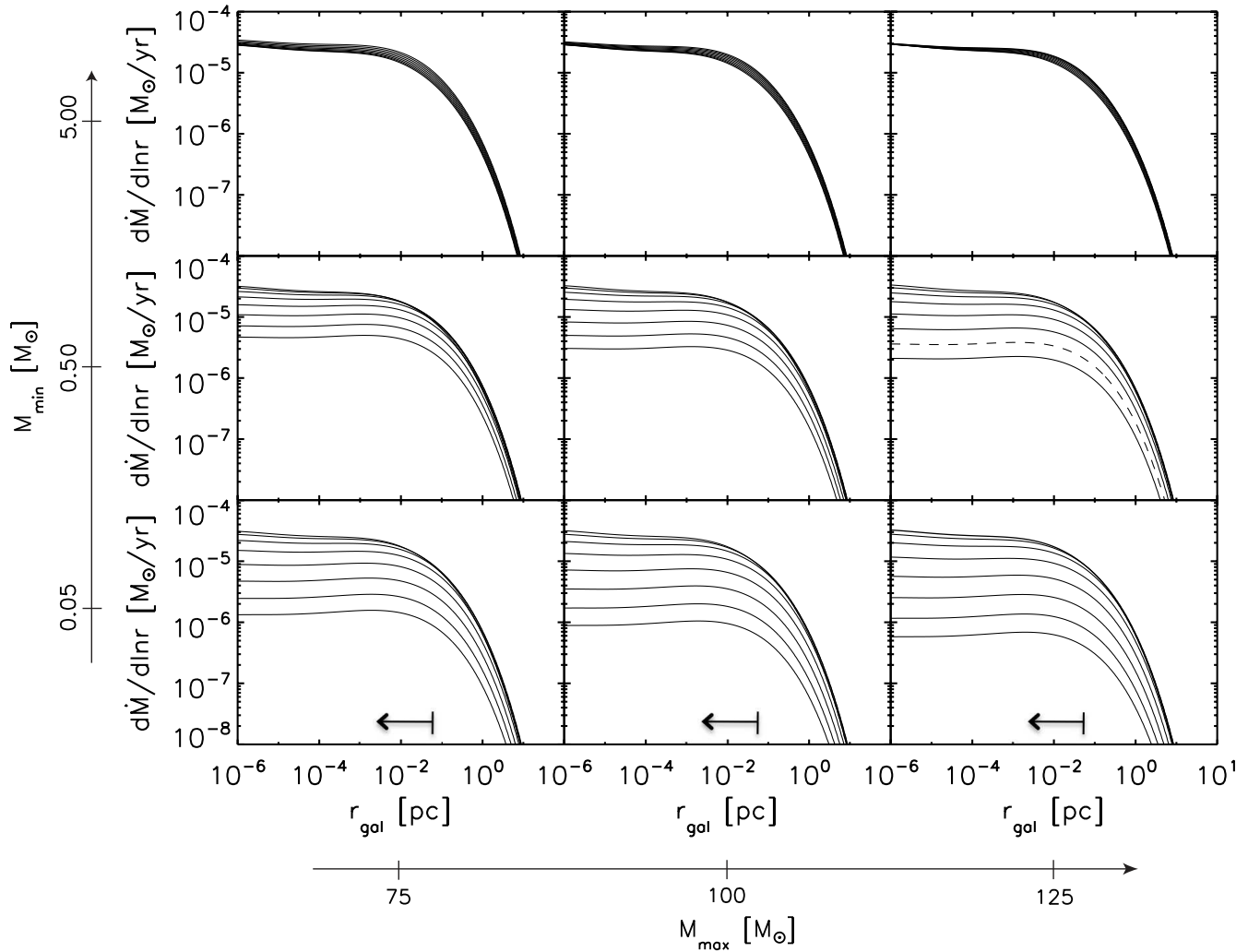}}
\caption{ \label{fig:dmdot_dlnr_v_r}Mass loss rates as a function of Galactic radius due to direct collisions for various parameters of $M_{min}$ $M_{max}$, and $\alpha$.  The parameter $M_{min}$ varies in each panel from bottom to top, and $M_{max}$ varies from left to right.  The power law slope, $\alpha$ varies within each panel from 1.00 (top line) to 2.5 (bottom line) in equal increments of 0.188.  The dashed line corresponds to a Salpeter-like mass function values ($M_{min} = 0.5 \mathrm{M_{\odot}}$, $M_{max} =125 \mathrm{M_{\odot}}$, $\alpha= 2.29$). The arrows indicate the range in the x-ray observations ($r_{gal} < 1.5''$) with which we use to constrain the PDMF (see \S \ref{sec:constrain}).}
\end{figure*}

Once values for $\alpha$, $M_{min}$ and $M_{max}$ are specified, equations~(\ref{eq:mass_loss_indir}) and~(\ref{eq:mass_loss_dir}) can be integrated to obtain the mass loss rate as a function of Galactic radius.  To show how the mass loss rate profiles vary with $M_{min}$, $M_{max}$ and $\alpha$, we plot $d\dot{M}/dlnr_{gal}$ for direct collisions in Fig.~\ref{fig:dmdot_dlnr_v_r} and vary these parameters.  In the figure, we have evaluated $M_{min}$ at 0.05, 0.5 and 5M$_{\odot}$, $M_{max}$ at 75, 100, 125M$_{\odot}$ and $\alpha$ from 1.00 to 2.5 in equal increments.  The parameter $M_{min}$ increases vertically from the bottom panel to the top, $M_{max}$ increases horizontally from the left panel to the right, and in each panel $\alpha$ increases from the bottom to the top.  We have indicated a Salpeter-like mass function ($\alpha =2.29$, $M_{min} =0.5M_{\odot}$ and $M_{max} =125M_{\odot}$) with the dashed line.  Mass loss is extensive and approximately constant until about $r_{gal}$ of $10^{-2}$pc and then drops dramatically.  This drop reflects that fact that collisions are less frequent at larger radii since star densities and relative velocities drop.  The amount of mass lost for any direct collision also decreases with galactic radius since $\Delta$ decreases with decreasing relative velocities.  Note that the profiles are approximately constant as a function of $M_{max}$, so that the choice of $M_{min}$ determines the extent of the mass loss rate.

\begin{figure}
\centerline{\includegraphics[clip, width=3.4in]{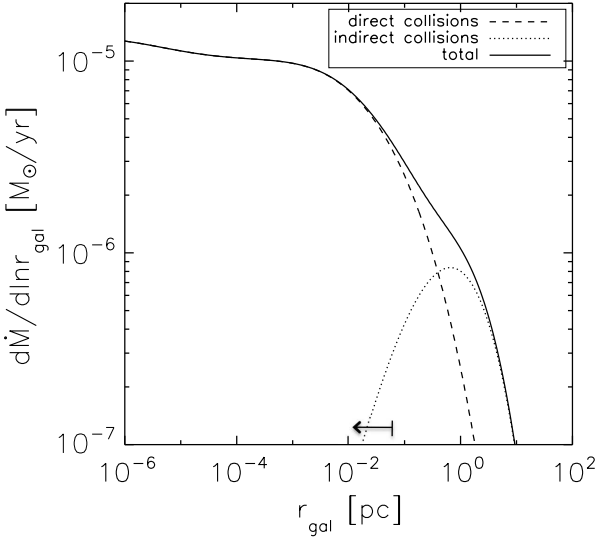}}
\caption{ \label{fig:dir_indir}Mass loss rates due to direct and indirect stellar collisions within the Galactic center for $M_{min}$=0.2M$_{\odot}$, $M_{max}$ = 100M$_{\odot}$ and $\alpha=1.2$.  The arrow indicates the range in the x-ray observations ($r_{gal} < 1.5''$) with which we use to constrain the PDMF (see \S \ref{sec:constrain}).}
\end{figure}

In Fig.~\ref{fig:dir_indir} we show the contributions to $d\dot{M}/dlnr_{gal}$ from both direct and indirect collisions for $M_{min}$=0.2M$_{\odot}$, $M_{max}$=100M$_{\odot}$ and $\alpha=1.2$.  We find that at small radii the mass loss rate is dominated by direct collisions, and at large radii it is dominated by indirect collisions.  Mass loss due to indirect collisions is suppressed in the Galactic center, due to the very fast relative stellar velocities.  Even though the high velocities (and high densities) in the Galactic center make collisions more frequent, under the impulse approximation, when velocities are very fast, mass loss is minimized.

\begin{figure*}
\centerline{\includegraphics[clip, width=7.1in]{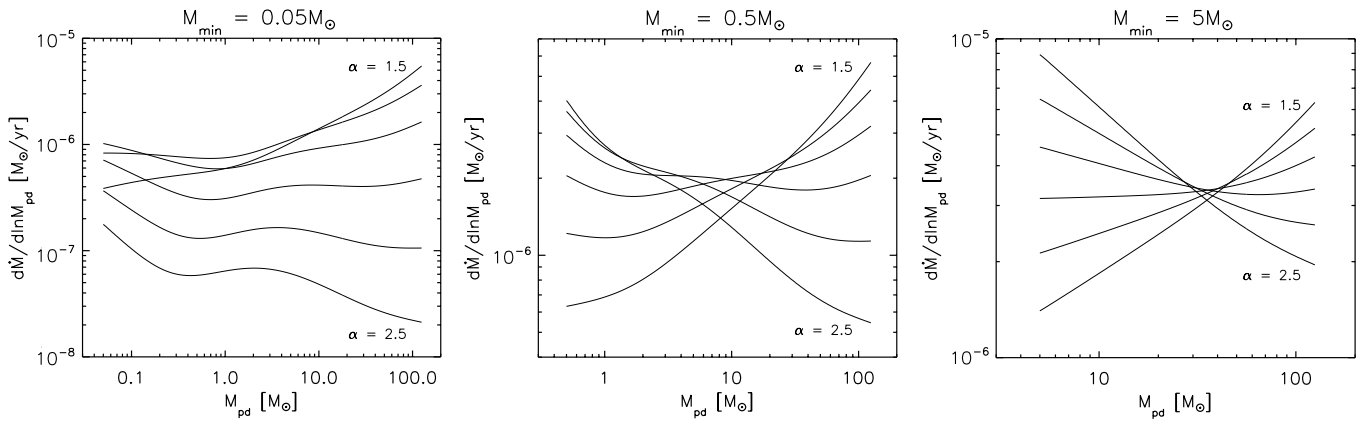}}
\caption{ \label{fig:dmdot_dlnm_v_m}  The amount of mass loss per logarithmic mass interval of the perturbed star as a function of the perturbed star's mass.   Each line was calculated with a different PDMF.  The titles in each panel indicate the value of $M_{min}$ used for that panel.  In each panel, $\alpha$ goes from 1.5 to 2.5 in even increments of 0.167, and for each line $M_{max} = 125 M_{\odot}$.}
\end{figure*}

To illustrate which mass stars contribute the most to the total mass loss rate, we plot $d\dot M/dlnM_{pd}$ as a function of $M_{pd}$ in Fig.~\ref{fig:dmdot_dlnm_v_m} for several different PDMFs.  The range of integration we choose for $r_{gal}$ is from 0 to 0.06pc (see \S \ref{sec:constrain}).  We choose $M_{min}$ to be 0.05, 0.5 and 5M$_{\odot}$(left to right in the figure), and we use a constant $M_{max}$ of 125M$_{\odot}$.  In each panel, we vary $\alpha$ from 1.5 to 2.5 in equal increments.  The figure shows that for $M_{min}=0.05$M$_{\odot}$, changing $\alpha$ has little effect on what mass stars contribute the most to the mass loss rate (although, the total mass loss rate is decreased with increasing $\alpha$).  For the $M_{min}=0.5$M$_{\odot}$ and $5$M$_{\odot}$ cases, increasing $\alpha$ results in lower mass stars contributing more to the mass loss rate.  This trend makes sense, since PDMF profiles with higher values of $\alpha$ have fractionally more lower mass stars.

To test how our interpolation between the $n=1.5$ and 3 polytropic indeces affects the main results of this paper, we consider two extreme cases.  The first case we consider has $n=1.5$ for $M_{\star}<1M_{\odot}$ and $M_{\star}>5M_{\odot}$, and $n=3$ for $1M_{\odot}\le M_{\star}\le 5M_{\odot}$.  This approach has $n=1.5$ for much of the mass spectrum, and should result in the highest mass loss rates since (as is evident from Fig.~\ref{fig:mass_loss}) collisions with the perturbed star having $n=1.5$ result in the most mass lost.  This is due to the fact that for $n=1.5$ stars, the mass is less centrally concentrated, and more mass can therefore escape at large radii which receive a stronger velocity kick.  The second case we consider has $n=1.5$ for $M_{\star}<0.3M_{\odot}$ and $M_{\star}>10M_{\odot}$, and $n=3$ for $0.3M_{\odot}\le M_{\star}\le 10M_{\odot}$.  This case should result in the smallest mass loss rates, since it has $n=1.5$ for a smaller fraction of the mass spectrum.  Since different mass functions have different fractions of the total mass in the neighborhood of $1M_{\odot}$ (where we expect the least mass loss per collision since $n=3$), we test the two cases for several different mass functions.  We find that differences in $d\dot{M}/dlnr_{gal} (r_{gal})$ for both cases are relatively minor, differing at most by $\sim10\%$ depending on the mass function that we use.

\section{Constraining the Mass Function in the Galactic Center}
\label{sec:constrain}

It is known through x-ray observations from \textit{Chandra}, that the central supermassive black hole in the Galactic center is surrounded by gas donated from stellar winds (e.g. Baganoff 2003).  The x-ray luminosity is due to Bremsstrahlung emission from unbound material supplied at a rate of $\sim 10^{-3} \mathrm{M_{\odot}yr^{-1}}$ \citep{najarro:1997}.  This unbound material has been studied theoretically by \citet{quataert:2004}, who solved the equations of hydrodynamics (under spherical symmetry) to follow how the gas is accreted onto Sgr A*.  \citet{quataert:2004} finds that his model agrees with the level of diffuse x-ray emission measured by \textit{Chandra}, and predicts an inflow of mass at $r_{gal} \sim 1''$ at a rate of $\sim 10^{-5} \mathrm{M_{\odot} yr^{-1}}$.

It is interesting to note that the mass loss rate profiles due to stellar collisions are consistent with the observed x-ray surface brightness profile.  From his derived temperature and density profiles, \citet{quataert:2004} solves for the x-ray surface brightness as a function of $r_{gal}$.  \citet{quataert:2004} has adopted a source term, $q(r_{gal})$ (the stellar mass loss rate per unit volume) of $q(r_{gal}) \propto r_{gal}^{-\eta}$ for $r_{gal} \in [2^{\prime \prime}, 10^{\prime \prime}]$, for $\eta = 0$, 2 and 3.  He finds that the model with $\eta = 3$ closely fits \textit{Chandra's} observed surface brightness profile.  From the definition of $q(r_{gal})$, $d\dot{M}/dlnr_{gal} \propto r_{gal}^3q(r_{gal}) \propto r_{gal}^{3-\eta}$.  It is evident from Fig.~\ref{fig:dmdot_dlnr_v_r} that in the range of $r_{gal}$ that \citet{quataert:2004} considers (about 0.08 to 0.39pc), $d\dot M/dlnr_{gal}$ is a power law.  Further, depending on which particular set of mass function parameters that we consider, $\eta \cong 2.6 - 2.8$ for stellar collisions.  These values of $\eta$ are very close to the model most consistent with the \textit{Chandra} surface brightness profile.

Using the the 2-10keV luminosity as measured by \textit{Chandra} \citep{baganoff:2003}, we estimate the total mass loss rate at a radius of $r_{gal} \sim 1.5''$ (0.06pc).  We use the word ``total'' to indicate the mass loss rate integrated over Galactic radius.  By using this total mass loss rate as an upper limit, we will be able to constrain the PDMF in the Galactic center by precluding any PDMFs with total mass loss rates greater than this value.  We will do this by integrating our calculated mass loss rate profiles (e.g., Figs.~\ref{fig:dmdot_dlnr_v_r} and~\ref{fig:dir_indir}) over $r_{gal}$.

Unbound material at a radius $r_{gal}$ has a dynamical timescale of
\begin{equation}
\label{eq:t_dyn}
t_{dyn}(r_{gal}) \sim \frac{r_{gal}}{v_{char}(r_{gal})} \approx 1.1 \times 10^{4} \mathrm{yrs} \left ( \frac{r_{gal}}{\mathrm{pc}}\right)^{1.5},
\end{equation}
where the characteristic velocity at radius $r_{gal}$, $v_{char}(r_{gal})$, is taken as the velocity dispersion as given in \S~\ref{sec:coll_rates}.   The electron density at radius $r_{gal}$ may therefore be estimated by:
\begin{align}
\label{eq:electron_dens}
n_e(r_{gal}) & \sim n_p(r_{gal})  \sim  \frac{\dot M t_{dyn}(r_{gal})}{\frac{4}{3} \pi r_{gal}^3 m_{p}} \nonumber \\ & =  1.1 \times 10^{5} \mathrm{cm^{-3}} \left (\frac{\dot M}{\mathrm{M_{\odot} yr^{-1}}} \right) \left ( \frac{r_{gal}}{\mathrm{pc}} \right)^{-1.5},
\end{align}
where $m_p$ is the proton mass.

For thermal Bremsstrahlung emission, the volume emissivity ($dE/dV dt d\nu$) is \citep{rybicki:1979}
\begin{align}
\label{eq:brem_form}
 \epsilon_{\nu}^{ff} &=  6.8 \times 10^{-38} \mathrm{erg\, s^{-1} cm^{-3} Hz^{-1}} \left(\frac{n_e}{\mathrm{cm^{-3}}}\right)^2 \left (\frac{T}{\mathrm{K}}\right)^{-1/2} \nonumber \\ & \times e^{-h \nu/k_B T} \bar{g}_{ff},
\end{align}
where we set $\bar g_{ff} =1.$  The luminosity in the 2-10keV band, $L_{2-10}$, can be found substituting equation~(\ref{eq:electron_dens}) into equation~(\ref{eq:brem_form}) and integrating the volume emissivity over volume (assuming spherical symmetry) and frequency:

\begin{align}
\label{eq:lum_2_10}
L_{2-10} & \sim  6.7 \times 10^{43} \mathrm{erg \, s^{-1}} \left (\frac{\dot M}{\mathrm{M_{\odot} yr^{-1}}} \right)^{2} \int_{r_{min}}^{0.06} \left (\frac{r}{\mathrm{pc}} \right)^{-1} d\left (\frac{r}{\mathrm{pc}}\right) \nonumber \\ & \times \int_2^{10} e^{-h \nu/\mathrm{keV}} d\left (\frac{h \nu}{\mathrm{keV}}\right).
\end{align}
We have assumed a constant temperature of 1keV.  A constant value of 1keV should suffice for an order of magnitude estimate as \citet{baganoff:2003} find that the gas temperature varies from approximately 1.9 to 1.3keV from $r_{gal}=$ 1.5 to 10'' (assuming an optically thin plasma model).  \citet{quataert:2004}'s model also predicts that the temperature varies from about 2.5 to 1keV from $r_{gal}=$ 0.3 to 10''.  By plugging the value of $L_{2-10}$ within 1.5'' ($2.4\times10^{33} \mathrm{erg \, s^{-1}}$) as measured by \citet{baganoff:2003} into equation~(\ref{eq:lum_2_10}) , we find $\dot M \sim 10^{-5} M_{\odot}yr^{-1}$.  This value is consistent with the mass inflow rate at $\sim 1''$ calculated by \citet{quataert:2004}.

Our results are not sensitive to the choice of the lower limit in the integral across $r_{gal}$.  The lower limit should be at most a few 0.01s of pcs to at least $\sim 10^{-6}$pc.  The former value is the tidal radius for the SMBH at the Galactic centre for a $1\mathrm{M_{\odot}}$ star.  Unbound material due to stellar collisions or from stellar wind should not exist at smaller radii since there are very few stars there to produce it.  The value of the integral thus ranges from about unity to a few 10s.  Since $\dot M$ depends upon the square root of this value, the exact value of $r_{min}$ only affects our calculation at the level of a factor of a few, and we thus take the square root of the integral to be unity.

\begin{figure}
\centerline{\includegraphics[clip, width=3.4in]{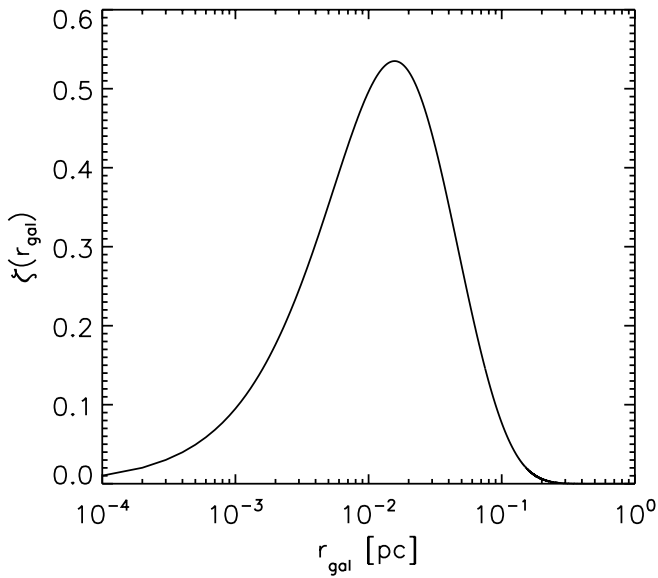}}
\caption{ \label{fig:zeta} The fraction of flux emitted from unbound material at radius $r_{gal}$ that contributes to the 2-10keV band.}
\end{figure}

Having established that $\dot M \sim 10^{-5} \mathrm{M_{\odot} yr^{-1}}$ in the vicinity of $1.5''$, we now calculate the expected mass loss rates due to stellar collisions for different PDMFs.  The value of $\dot M$ that contributes to the 2-10keV flux is given by:
\begin{equation}
\label{eq:total_mass_loss_rate}
\dot M = \int_{0}^{0.06\mathrm{pc}} \frac{d \dot M}{d^3r_{gal}} \zeta(r_{ral}) d^3 r_{gal},
\end{equation}
where we have shown how to calculate the mass loss rate profiles, $d \dot M/d^3r_{gal}$ in the previous section.  We account for the fact that not all of the emission from the unbound gas contributes to the 2-10keV band with $\zeta(r_{gal})$, defined as the fraction of flux from gas at radius $r_{gal}$ with $2\mathrm{keV}  \le h\nu \le 10\mathrm{keV}$:
\begin{equation}
\label{eq:zeta}
\zeta(r_{gal}) \equiv \frac{\int_{2keV}^{10keV}\epsilon_{\nu}^{ff} d\nu}{\int_{0}^{\infty} \epsilon_{\nu}^{ff}d \nu} = e^{-2keV/k_B T(r_{gal})} - e^{-10keV/k_B T(r_{gal})}.
\end{equation}
Since the gas at each radius is at a slightly different temperature, and since $\zeta$ is exponentially sensitive to the temperature, we must estimate $T(r_{gal})$.  We do this by setting the thermal energy of the unbound material equal to the kinetic energy at a radius $r_{gal}$, and find that
\begin{equation}
\label{eq:temp}
k_B T(r_{gal}) \approx m_p \sigma^2(r_{gal})  = 7.8 \times 10^{-2} \mathrm{keV} \left(\frac{r_{gal}}{pc}\right)^{-1}.
\end{equation}
We plot equation~(\ref{eq:zeta}) in Fig.~\ref{fig:zeta}.  The value of $\zeta$ goes to zero at the highest and smallest radii since, for the former, the gas is cool and emits most of its radiation redward of 2keV, and for the latter, the gas is hot and emits mostly blueward of 10keV.  Thus, even though the integral in equation~(\ref{eq:total_mass_loss_rate}) extends to $r_{gal} = 0$, the contribution to $\dot M$ is suppressed exponentially at the smallest radii.

Since, by equations (\ref{eq:mass_loss_indir}) and (\ref{eq:mass_loss_dir}), $\dot{M}$ depends on the parameters of the PDMF, we now constrain these parameters by limiting the allowed mass loss rate from stellar collisions calculated via equation~(\ref{eq:total_mass_loss_rate}) at $10^{-5} \mathrm{M_{\odot} yr^{-1}}$.  We consider changes in $M_{min}$ and $\alpha$, and keep $M_{max}$ set at $125\mathrm{M_{\odot}}$ since (as seen in Fig.~\ref{fig:dmdot_dlnr_v_r}) $\dot{M}$ is approximately independent of $M_{max}$.

\begin{figure}
\centerline{\includegraphics[clip, height=3.6in, angle=90]{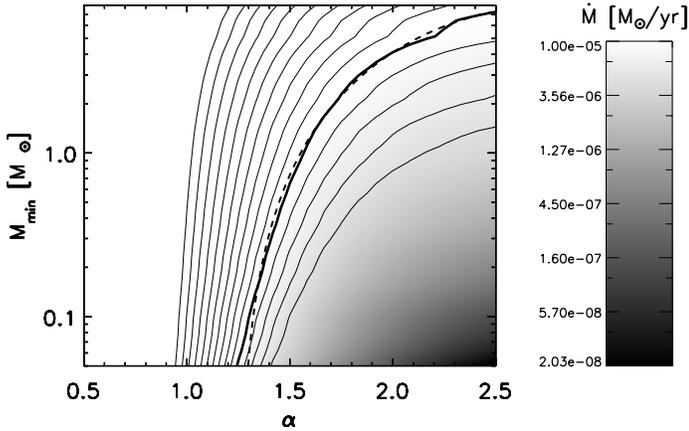}}
\caption{ \label{fig:param_space} The total mass loss rate contributing the 2-10keV flux (calculated from equation~(\ref{eq:total_mass_loss_rate})) as a function of $M_{min}$ and $\alpha$. The solid contours are on a logarithmic scale, and are limited from above at $10^{-5}\mathrm{M_{\odot}yr^{-1}}$. The line contours are on a linear scale, and are separated by intervals of $1.5 \times 10^{-6}$M$_{\odot}$yr$^{-1}$.  The thick line denotes the $1 \times 10^{-5}$M$_{\odot}$yr$^{-1}$contour.  The dashed line is a 3rd order polynomial fit which represents the absolute allowed upper limit of $M_{min}$ as a function of $\alpha$.  The coefficients of this polynomial are 21.71, -42.37, 25.33 and -4.27 for $a_0$ to $a_3$ respectively.}
\end{figure}

We sample the $M_{min}- \alpha$ parameter space and use equation~(\ref{eq:total_mass_loss_rate}) to compute the total mass loss rate, the results of which are shown in Fig.~\ref{fig:param_space}.  The contours represent the calculated $\dot{M}$ values, where the solid contours are on a logarithmic scale, and where they are limited from above at a value of $\dot{M} = 10^{-5}\mathrm{M_{\odot}yr^{-1}}$.  The lines are on a linear scale with intervals of $1.5 \times 10^{-6}$M$_{\odot}$yr$^{-1}$, with the thick line denoting the $10^{-5}$M$_{\odot}$yr$^{-1}$ level. The figure shows that PDMFs with flat to canonical-like profiles are allowed.  Very top-heavy profiles ($\alpha \lesssim 1.25$) are not allowed, as they predict too high of a mass loss rate.  Mass functions with $M_{min} \gtrsim 7$M$_{\odot}$ are also not allowed.  These results are consistent with measurements of the Arches star cluster, a young cluster located about 25pc from the Galactic center.  Recent measurements \citep{kim:2006, stolte:2005, figer:1999} probing stellar masses down to about $1\mathrm{M_{\odot}}$ show that the cluster has a flat PDMF, with $\alpha$ in the range of about 1.2 to 1.9 (depending on the location within the cluster).

Since $\dot M$ is a much stronger function of $\alpha$ than of $M_{min}$ it is difficult for us to place tight constraints on the allowed range of $M_{min}$.  Fig.~\ref{fig:param_space} shows that we can, however, place a constraint on the allowed upper limit of $M_{min}$, since very high values of $M_{min}$ result mass loss rates $> 10^{-5}\mathrm{M_{\odot}yr^{-1}}$.  For $\alpha > 1.25$, we fit a 3rd degree polynomial (the dashed line in Fig.~\ref{fig:param_space}) to the $\dot M = 10^{-5}\mathrm{M_{\odot} yr^{-1}}$ contour.  This fit analytically expresses the upper limit of $M_{min}$ as a function of $\alpha$.  We provide the coefficients of this fit in the caption of Fig.~\ref{fig:param_space}.

The small difference between the solid and dashed lines at $r_{gal} = 0.06$pc in Fig.~\ref{fig:diff_coll_rate}b suggests that, even for stellar encounters involving small impact parameters, our integration does not miss many collisions by ignoring gravitational focusing.  To estimate the contribution to the total mass loss rate in Fig.~\ref{fig:param_space} from gravitational focusing, we take $\delta M_{typ}$, the typical amount of mass lost per collision, to be simply a function of $b$.  This avoids the multi-dimensional integrations involved in equations (\ref{eq:mass_loss_indir} and~\ref{eq:mass_loss_dir}), since for these equations $\Delta_{pd}$ is a function of $b$, $M_{pr}$, $M_{pd}$, and $v_{rel}(r_{gal})$.  For simplicity, we choose $\delta M_{typ} (b)$ to decrease linearly from 2M$_{\odot}$ (we assume that both stars are completely destroyed) at $b=0$ to 0 at $b=b_0$.  We find $b_0$ by noting from Fig.~\ref{fig:mass_loss} that for all values of the polytropic index, the amount of mass loss for an indirect collision goes to zero at around $\gamma =0.98$.  By recalling the definition of $\gamma$ (equation (\ref{eq:def_of_gamma})), we solve for $b_0$ at $\gamma =0.98$ by setting $\tilde M_{pr} =1$, and taking $v_{rel} \sim 2\sigma(r_{gal}=0.06 \mathrm{pc})$.  By calculating $d\Gamma/db~( < r_{gal})$ (for Salpeter values) evaluated at 0.06pc across a range of $b$, and multiplying by $\delta M_{typ} (b)$, we are able to estimate $d \dot M /db$.  We do this for $d \Gamma/db~( < r_{gal})$ with and without gravitational focusing, and integrate across $b$.  Subtracting the two numbers results in our estimate of the contribution to the total mass loss rate due to gravitational focusing: $2.3 \times 10^{-7}$M$_{\odot}$.  This is about twice the mass loss rate from Fig.~\ref{fig:param_space} evaluated at Salpeter values.  We perform the same calculation across the $M_{min}$ - $\alpha$ parameter space, and find that gravitational focusing contributes a factor of at most $\sim 2.5$ to the total mass loss rate.

An underestimate of a factor of 2.5 slightly affects the region of parameter space that we are able to rule out, as shown by the line contours in Fig.~\ref{fig:param_space}.  The contours are on a linear scale, starting at $4 \times 10^{-6}$M$_{\odot}$yr$^{-1}$, and ending at $2.5 \times 10^{-5}$M$_{\odot}$yr$^{-1}$, in intervals of $1.5 \times 10^{-6}$M$_{\odot}$yr$^{-1}$.  The $4 \times 10^{-6}$M$_{\odot}$yr$^{-1}$ contour (2.5 times less than the $10^{-5}$ contour) shows that the region of the parameter space that is ruled out is $M_{min} \gtrsim 1.4$M$_{\odot}$ and $\alpha \lesssim 1.4$.

As noted in \S \ref{sec:mass_loss_direct}, the amount of mass loss that we calculate for a single direct collision sometimes over or under-predicts the amount of mass lost (as compared to the work of \citet{freitag:2005}) by a factor of a few to at most a factor of about 10 (see Figs.~\ref{fig:comp_1} and~\ref{fig:comp_2}).  We find that our calculation of mass loss for direct collisions shows no preference over whether the amount of mass lost is over, or under-predicted when considering different combinations of $\tilde M_1$, $\tilde M_2$, $\tilde v_{rel}$ and $\tilde b$.  We therefore suspect that when integrating over all of these parameters to obtain the total amount of mass lost, our error will roughly cancel.  The line contours in Fig.~\ref{fig:param_space} serve as a good gauge of how our constraints on the allowed region of the parameter change given an uncertainty of a factor of a few in our calculation of $\dot M$.  As noted in the previous paragraph, an underestimation of the total mass loss rate results in too high of an allowed upper limit in $M_{min}$ by a factor of a few, and too low of an allowed lower limit in $\alpha$ by about 10\%.  An overestimation of the total mass loss rate of a factor of a few results in too high of an allowed lower limit in $\alpha$ by a few tens of percents.  An overestimation will also result in a significantly too low allowed upper limit in $M_{min}$ since (as seen in the figure) $\dot M$ is a much stronger function of $\alpha$ for small $\alpha$ so that the contours are nearly vertical.

\section{Implications for the IMF}

We now place constraints on the IMF in the Galactic centre with a simple analytical approach that connects the IMF to the PDMF, and with the results of the previous section.  The mass function as a function of time is described by a partial differential equation that takes into account the birth rate and death rate of stars:
\begin{equation}
\label{eq:diff_q}
\frac{\partial \xi (M, t)}{\partial t} = R_B(t)\Phi(M) -\xi \frac{1}{\tau_{\star}(M)},
\end{equation}
where $R_B (t)$ is the birth rate density of stars ($dN_B/(dt d^3r_{gal})$), $\Phi(M)$ is the initial mass function normalized such that  $\int \Phi(M) dM =1$, and $\tau_\star (M)$ is the main sequence lifetime of stars as a function of stellar mass.  For the initial mass function, we take a power law,
\begin{equation}
\label{eq:IMF}
\Phi = M^{-\gamma},
\end{equation}
and for $\tau_{\star}(M)$ we use the expression given by \citet{mo:2010}

\begin{equation}
\tau_\star(M) = \frac{2.5 \times 10^3 + 6.7 \times 10^2 M^{2.5} +M^{4.5}}{3.3 \times 10^{-2} M^{1.5} +3.5 \times 10^{-1} M^{4.5}} \mathrm{Myr},
\end{equation}
valid for $0.08$M$_\odot< M< 100$M$_\odot$ and for solar-type metallicity.

In the following paragraphs, we consider different star formation history scenarios.  For each scenario, we will need to know $R_B(\tau_{MW})$, the star formation rate density in the Galactic centre at the age of the Milky Way (which we take to be 13Gyr).  A rough estimate of this value is given by the number density of young stars in the Galactic centre divided by their age: $R_B(\tau_{MW}) \sim \bar \rho (r) \eta/ (\langle \tau \rangle \langle M \rangle)$.  Here $\langle \tau \rangle$ and $\langle M \rangle$ are the average age and average mass of the young stars in the Galactic center, which we take to be $\sim 10$Myr and $\sim 10$M$_\odot$ respectively.  The parameter $\eta$ is the fraction of stars with masses above $10$M$_\odot$, which for reasonable mass functions is $\sim 0.1$\%.  For self-consistency, we use $\bar \rho$ evaluated at $0.06$pc (which from equation (\ref{eq:rho_enc}) is $\sim 10^7$M$_\odot$pc$^{-3}$), since this was the radius with which we used to constrain the present-day mass function.  These values result in $R_{B}(\tau_{MW}) \sim 10^{-4} $pc$^{-3}$yr$^{-1}$.

For the simple case of a constant star formation rate, $R_{B}(t) = R_B(\tau_{MW})$, and the solution to equation (\ref{eq:diff_q}) with the boundary condition that $\xi(M, t=0) = \Phi(M)n_{tot}(t=0)$, evaluated at the current age of the Milky-way is:
 \begin{align}
 \label{eq:PDMF_solution}
 \xi(M, t=\tau_{MW}) & =\Phi(M)e^{-\tau_{MW}/\tau_\star(M)} \nonumber \\
 &\times \left \{ R_B(\tau_{MW}) \tau_\star(M)e^{\tau_{MW}/\tau_\star(M)} \right. \nonumber \\
&  -R_B(\tau_{MW}) \tau_\star(M) +n_{tot}(0) \left. \vphantom{ R_B(\tau_{MW}) \tau_\star(M)e^{\tau_{MW}/\tau_\star(M)} } \right \}.
 \end{align}
 We evaluate the solution at the age of the Milky-way (yielding the PDMF) because want to compare with our constraints on the PDMF as found in the previous section.  To solve for $n_{tot}(0)$, we use the known mean density of the Galactic centre today at 0.06pc, $\bar \rho(\tau_{MW}, r=0.06\mathrm{pc})$, insert equation (\ref{eq:PDMF_solution}) into the following expression:
 \begin{equation}
 \label{eq:no_eqn}
\bar \rho(\tau_{MW}, r=0.06\mathrm{pc}) = \int \xi(M, t=\tau_{MW})MdM,
\end{equation}
and solve for $n_{tot}(0)$.

\begin{figure}
\centerline{\includegraphics[clip, height=1.5in]{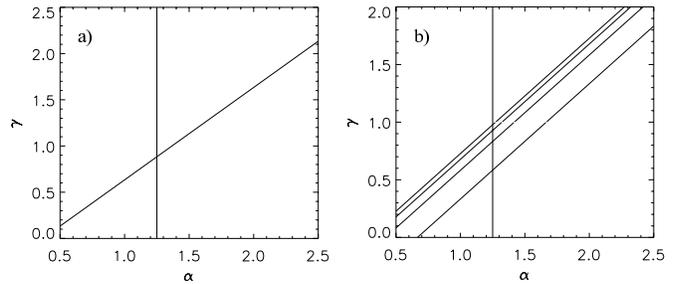}}
\caption{ \label{fig:const_exp} a) The IMF power-law slope as a function of the PDMF power-law slope for the case of constant star formation.  b) The same except for exponentially decreasing star formation with $\tau_{exp}=$ 3, 5, 7, 9Gyr (bottom to top line respectively).}
\end{figure}

We solve for $\xi(M, \tau_{MW})$ for a range of different IMF power-law slopes, $\gamma$, and fit a power-law to the solution, with a power-law slope $\alpha$.  We plot the IMF power-law slope as a function of the calculated PDMF power-law slope for constant star formation in Fig.~\ref{fig:const_exp}a. We have constrained the PDMF in the previous section to have $\alpha \gtrsim 1.25$, indicated in the figure by the vertical line.  The figure therefore shows that for the case of constant star formation, the IMF power-law slope, $\gamma$, must be $\gtrsim 0.9$.

For the general case of a star formation rate that varies with time, $R_B(t) \neq R_B(\tau_{MW})$, and the solution to equation (\ref{eq:diff_q}) with the same boundary condition and evaluated at $\tau_{MW}$ is:
\begin{align}
\label{eq:xi_general}
 \xi(M, t=\tau_{MW}) & =\Phi(M)e^{-\tau_{MW}/\tau_\star(M)} \\ \nonumber
&\times  \left \{ \int_0^{\tau_{MW}} R_B(t^\prime) e^{t^{\prime}/\tau_\star(M)}dt^\prime +n_{tot}(0) \right \}.
\end{align}
For an exponentially decreasing star formation history, the star formation rate is given by:
\begin{equation}
R_B(t) = R_B(\tau_{MW}) e^{-(t-\tau_{MW})/\tau_{exp}}.
\end{equation}
Given this star formation history, we solve for $\xi(M, \tau_{MW})$ (by solving for $n_{tot}(0)$ with equation (\ref{eq:no_eqn})) for $\tau_{exp} = $3, 5, 7 and 9Gyr. We fit power-laws to the resulting PDMFs, and show the results in Fig.~\ref{fig:const_exp}b.  The figure shows that smaller values of $\tau_{exp}$ result in larger values of $\alpha$ for any given $\gamma$.  The trend can be explained by the fact that since a smaller value of $\tau_{exp}$ results in a steeper $R_B$ profile, and that all profiles must converge to $R_B(\tau_{MW})$ at the present-time, $R_B$ profiles with smaller values of $\tau_{exp}$ have had overall more star formation in the past.  More overall star formation means that the present-day mass function is comprised of fractionally more lower-mass stars since the IMF favors lower-mass stars.  The constant build-up of lower-mass stars results in a steeper PDMF, so that for any given $\gamma$, $\alpha$ should be larger.  The figure shows that for exponentially decreasing star formation $\gamma$ must be $\gtrsim$ 0.6, 0.8, 0.9 and 1.0 for $\tau_{exp} = $ 3, 5, 7, and 9Gyr respectively.

\begin{figure}
\centerline{\includegraphics[clip, height=3.3in]{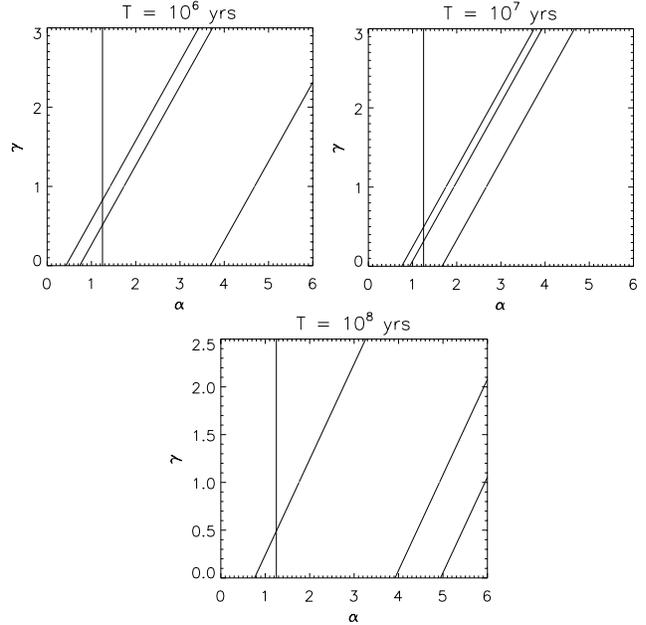}}
\caption{ \label{fig:step} The IMF power-law slope as a function of the PDMF power-law slope for the case of episodic star formation.  In each panel the lowest line is $\Delta t = 10^8$yrs and the highest line is $\Delta t = 10^6$yrs.}
\end{figure}

 The final case we consider is episodic star formation, where each episode lasts for a duration $\Delta t$, where the ending and beginning of each episode is separated by a time, $T$, and where the magnitude of each episode is $R_B(\tau_{MW})$.  For such a star formation history, the solution to equation (\ref{eq:xi_general}) is:

\begin{align}
& \xi(M, t=\tau_{MW})  =\Phi(M)e^{-\tau_{MW}/\tau_\star(M)} \nonumber \\
&\times  \left \{R_B(\tau_{MW}) \tau_\star(M) \sum_{n=0}^{n_{max}} \left [e^{[(n+1)\Delta t +nT]/\tau_\star(M)} \right. \right. \\
& - e^{n(\Delta t +T)/\tau_\star(M)}\left. \vphantom{e^{[(n+1)\Delta t +nT]/\tau_\star(M)}} \right ] +n_{tot}(0) \left. \vphantom{\sum_{n=0}^{n_{max}}} \right \},
\end{align}
where $n_{max}=\mathrm{floor}\{ (\tau_{MW}-\Delta t)/(T+\Delta t)\}$, and where we again solve for $n_{tot}(0)$ with equation (\ref{eq:no_eqn}).  We consider 9 cases with $\Delta t$ and $T = 10^6$, $10^7$, and $10^8$yrs, and show the results in Fig.~\ref{fig:step}.  In each panel the lowest line is $\Delta t = 10^8$yrs and the highest line is $\Delta t = 10^6$yrs.  For $T = 10^6$yrs, $\gamma \gtrsim $ 0.8 and 0.5 for $\Delta t = 10^6$ and $10^7$yrs respectively, while the $\Delta t = 10^8$yrs case results in constraints on $\gamma$ that are too low to be realistic.  For $T = 10^7$yrs, $\gamma \gtrsim $ 0.5 and 0.4 for $\Delta t = 10^6$ and $10^7$yrs respectively, while again, the $\Delta t = 10^8$yrs case results in unrealistic constraints.  Finally, for the $T = 10^8$yrs, $\gamma \gtrsim $ 0.5 for $\Delta t = 10^6$, while the $\Delta t = 10^7$ and $10^8$yrs case result in unrealistic constraints.  We test whether when the last star formation episode occurs relative to the present day affects our solution of $\xi(M, \tau_{MW})$ by varying the start time of the star formation episodes.  By varying the start time and testing all the combinations of $\Delta t$ and $T$ that we consider, we find that the lines in Fig.~\ref{fig:step} vary by at most about 5\%, so that the main trends in the figure are unaffected.

\section{Contribution from red giants}

Spectroscopic observations have revealed that the central parsec of the Galaxy harbors a significant population of giant stars \citep{paumard:2006, maness:2007}.  Due to their large radii (and hence large cross sections), it is possible that they could play an important part in the mass loss rate due to collisions in the Galactic centre.

In assessing their contribution to the mass loss rate, care must be taken when deriving the collision rates, because their radii, $r_{RG}$, are strong functions of time, $t$.  \citet{dale:2009} have already calculated the probability, $P(r_{gal})$, for a red giant (RG) in the Galactic centre to undergo collisions with main sequence impactors.  They have taken into account that $r_{RG} (t)$ by integrating the collision probability over the time that the star resides on the RG branch.  We use their results to estimate the mass loss rate due to RG - MS star collisions.

To find the number density of RGs in the Galactic centre, we weight the total stellar density by the fraction of time the star spends on the RG branch:
\begin{equation}
\label{eqn:n_rg}
n_{RG}(r_{gal}) \sim n_{\star}(r_{gal})\frac{\tau_{RG}}{\tau_{\star}}.
\end{equation}
This approximation should be valid given a star formation history that is approximately constant when averaged over time periods of order $\tau_{RG}$.  The number of collisions per unit time suffered by any one red giant, $\dot P(r_{gal})$, should of order the collision rate averaged over the lifetime of the RG, and is given by
\begin{equation}
\dot P(r_{gal}) \sim \left \langle \dot P(r_{gal}) \right \rangle_{t} = \frac{P(r_{gal})}{\tau_{RG}}.
\end{equation}
If we define $\delta M$ to be the typical amount of mass lost in the collision, then the mass loss rate is
\begin{equation}
\label{eqn:mdot_rg}
\frac{d \dot M}{dlnr_{gal}} = 4 \pi r_{gal}^3 \frac{d \dot M}{d^3r_{gal}} \sim 4 \pi r_{gal}^3 n_{RG}(r_{gal}) \frac{P(r_{gal})}{\tau_{RG}} \delta M.
\end{equation}

To calculate an upper limit for the contribution of RG - MS star collisions to the mass loss rate, we assume all RG and MS stars have masses of $1\mathrm{ M_{\odot}}$, and that the entire RG is destroyed in the collision.  Collisions involving $1\mathrm{M_{\odot}}$ RGs yield an upper limit, because there is not an appreciable amount of RGs with masses less than $\sim 1\mathrm{M_{\odot}}$ due to their MS lifetimes being greater than the age of the Galaxy.  For RGs with masses greater than $1 \mathrm{M_{\odot}}$, the amount they contribute to the mass loss rate is a competition between their lifetimes and radii.  Red giant lifetimes decrease with mass (thereby decreasing the time they have to collide) and their radii increase with mass (thereby increasing the cross section).  In their Fig. 3, \citet{dale:2009} clearly show that the number of collisions decreases with increasing RG mass, indicating that the brevity of their lifetime wins over their large sizes.  One solar mass MS impactors should yield approximately an upper limit to the mass loss rate, since $\sim 1\mathrm{M_{\odot}}$ MS stars are the most common for the PDMFs under consideration.

\begin{figure}
\centerline{\includegraphics[clip, width=3.in]{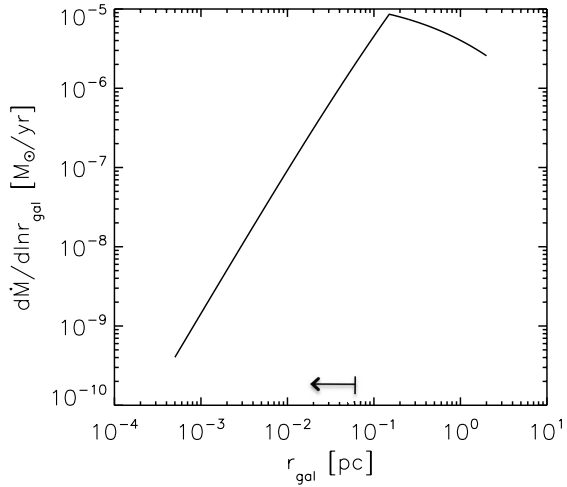}}
\caption{ \label{fig:red_giant_cont}  An upper limit to the mass loss rate due to collisions between RG and MS stars. The arrow indicates the range in the x-ray observations ($r_{gal} < 1.5''$) with which we use to constrain the PDMF (see \S \ref{sec:constrain}).}
\end{figure}

Since we assume that the entire RG is destroyed in the collision $\delta M = 1\mathrm{M_{\odot}}$.  For the case that all impactors are $1\mathrm{M_{\odot}}$ MS stars, we calculate $n_{RG} (r_{gal})$ from equation (\ref{eqn:n_rg}) by noting that $n_{\star}(r_{gal}) = \rho_{\star}(r_{gal})/(1M_{\odot})$.  For self-consistency, we must truncate $P(r_{gal})$ at 1 for all $P(r_{gal}) >1$ since we are considering the case where one collision destroys the entire star.  We plot equation (\ref{eqn:mdot_rg}) for this calculation in Fig.~\ref{fig:red_giant_cont}.  The discontinuity is due to our truncating $P(r_{gal})$ at 1.  The figure shows that the mass loss rate for RG-MS star collisions never exceeds $10^{-5}\mathrm{M_{\odot}yr^{-1}}$, well below typical $d\dot M/dlnr_{gal}$ for values for MS - MS collisions (see Figs.~\ref{fig:dmdot_dlnr_v_r} and~\ref{fig:dir_indir}).  Moreover, in their hydrodynamic simulations, \citet{dale:2009} note that in a typical RG - MS star collision, at most $\sim 10\%$ of the RG envelope is lost to the RG.  We therefore conclude that the contribution of RGs to the total mass loss rate in the central parsec of the Galaxy is negligible.

The figure shows that by $r_{gal} = 0.06$pc, the mass loss rate for RG-MS star collisions is at most about $10^{-6}\mathrm{M_{\odot}yr^{-1}}$.  It is thus possible that for MS-MS collisions, values of $M_{min}$ and $\alpha$ that results in total mass loss rates just below $10^{-5}\mathrm{M_{\odot}yr^{-1}}$ could be pushed past this threshold with the addition of mass loss due to RG collisions.  However, we believe that this is unlikely for two reasons.  The inclusion of the factor, $\zeta$, when calculating the total mass loss rate (see equation~(\ref{eq:total_mass_loss_rate})) will reduce the mass loss by at least a factor of 0.6 (see Fig.~\ref{fig:zeta}).  Also, as noted by the hydrodynamic simulations of \citet{dale:2009}, for a typical RG - MS star collision, at most $\sim 10\%$ of the RG envelope is lost to the RG.  This will reduce $d \dot M /dlnr_{gal}$ for RG-MS collisions by another factor of 10.

\section{Conclusions}

We have have derived novel, analytical methods for calculating the amount of mass loss from indirect and direct stellar collisions in the Galactic centre.  Our methods compares very well to hydrodynamic simulations, and do not require costly amounts of computation time.  We have also computed the total mass loss rate in the Galactic centre due to stellar collisions.  Mass loss from direct collisions dominates at Galactic radii below $\sim0.1\mathrm{pc}$, and thereafter indirect collisions dominate the total mass loss rate.  Since the amount of stellar material lost in the collision depends upon the masses of the colliding stars, the total mass loss rate depends upon the PDMF.  We find that the calculated mass loss rate is sensitive to the PDMF used, and can therefore be used to constrain the PDMF in the Galactic centre.  As summarized by Fig.~\ref{fig:param_space}, our calculations rule out $\alpha \lesssim 1.25$ and $M_{\min} \gtrsim 7 \mathrm{M_{\odot}}$ in the $M_{min}-\alpha$ parameter space.  Finally, we have used our constraints on the PDMF in the Galactic centre to constrain the IMF to have a power-law slope $\gtrsim $ 0.4 to 0.9 depending on the star formation history of the Galactic centre.

\section*{Acknowledgments}

This work was supported in part by the National Science Foundation Graduate Research Fellowship, NSF grant AST-0907890 and NASA grants NNX08AL43G and NNA09DB30A.

\label{lastpage}


\begin{thebibliography}{99}

\bibitem[\protect\citeauthoryear{Aguilar
\& White}{1985}]{aguilar:1985} Aguilar L.~A., White S.~D.~M., 1985, ApJ, 295, 374

\bibitem[\protect\citeauthoryear{Alexander}{1999}]{alexander:1999}
Alexander T., 1999, ApJ, 527, 835

\bibitem[\protect\citeauthoryear{Baganoff et
al.}{2003}]{baganoff:2003} Baganoff F.~K., et al., 2003, ApJ, 591,
891

\bibitem[\protect\citeauthoryear{Bailey
\& Davies}{1999}]{bailey:1999} Bailey V.~C., Davies M.~B., 1999, MNRAS, 308, 257

\bibitem[\protect\citeauthoryear{Bartko et al.}{2010}]{bartko:2010}
Bartko H., et al., 2010, ApJ, 708, 834

\bibitem[\protect\citeauthoryear{Bastian, Covey,
\& Meyer}{2010}]{bastian:2010} Bastian N., Covey K.~R., Meyer M.~R., 2010, ARA\&A, 48, 339

\bibitem[\protect\citeauthoryear{Benz
\& Hills}{1987}]{benz:1987} Benz W., Hills J.~G., 1987, ApJ, 323, 614

\bibitem[\protect\citeauthoryear{Benz
\& Hills}{1992}]{benz:1992} Benz W., Hills J.~G., 1992, ApJ, 389, 546

\bibitem[\protect\citeauthoryear{Binney
\& Tremaine}{2008}]{binney:2008} Binney J., Tremaine S., 2008, gady.book,

\bibitem[\protect\citeauthoryear{Dale et al.}{2009}]{dale:2009}
Dale J.~E., Davies M.~B., Church R.~P., Freitag M., 2009, MNRAS, 393, 1016

\bibitem[\protect\citeauthoryear{Davies et al.}{1998}]{davies:1998}
Davies M.~B., Blackwell R., Bailey V.~C., Sigurdsson S., 1998, MNRAS, 301,
745

\bibitem[\protect\citeauthoryear{DeYoung}{1968}]{deyoung:1968} DeYoung
D.~S., 1968, ApJ, 153, 633

\bibitem[\protect\citeauthoryear{Eckart et al.}{1993}]{eckart:1993}
Eckart A., Genzel R., Hofmann R., Sams B.~J., Tacconi-Garman L.~E., 1993,
ApJ, 407, L77

\bibitem[\protect\citeauthoryear{Eckart et al.}{2002}]{eckart:2002}
Eckart A., Genzel R., Ott T., Sch{\"o}del R., 2002, MNRAS, 331, 917

\bibitem[\protect\citeauthoryear{Figer et al.}{1999}]{figer:1999}
Figer D.~F., Kim S.~S., Morris M., Serabyn E., Rich R.~M., McLean I.~S.,
1999, ApJ, 525, 750

\bibitem[\protect\citeauthoryear{Freitag
\& Benz}{2005}]{freitag:2005} Freitag M., Benz W., 2005, MNRAS, 358, 1133

\bibitem[\protect\citeauthoryear{Genzel et al.}{1996}]{genzel:1996}
Genzel R., Thatte N., Krabbe A., Kroker H., Tacconi-Garman L.~E., 1996,
ApJ, 472, 153

\bibitem[\protect\citeauthoryear{Genzel et al.}{2003}]{genzel:2003}
Genzel R., et al., 2003, ApJ, 594, 812

\bibitem[\protect\citeauthoryear{Ghez et al.}{2003}]{ghez:2003}
Ghez A.~M., et al., 2003, ApJ, 586, L127

\bibitem[\protect\citeauthoryear{Ghez et al.}{2008}]{ghez:2008}
Ghez A.~M., et al., 2008, ApJ, 689, 1044

\bibitem[\protect\citeauthoryear{Gnedin, Hernquist,
\& Ostriker}{1999}]{gnedin:1999} Gnedin O.~Y., Hernquist L., Ostriker J.~P., 1999, ApJ, 514, 109

\bibitem[\protect\citeauthoryear{Kim et al.}{2006}]{kim:2006}
Kim S.~S., Figer D.~F., Kudritzki R.~P., Najarro F., 2006, ApJ, 653, L113

\bibitem[\protect\citeauthoryear{Kippenhahn
\& Weigert}{1994}]{kippenhahn:1994} Kippenhahn R., Weigert A., 1994, sse..book,

\bibitem[\protect\citeauthoryear{Kroupa}{2001}]{kroupa:2001} Kroupa
P., 2001, MNRAS, 322, 231

\bibitem[\protect\citeauthoryear{Lai, Rasio,
\& Shapiro}{1993}]{lai:1993} Lai D., Rasio F.~A., Shapiro S.~L., 1993, ApJ, 412, 593

\bibitem[\protect\citeauthoryear{L{\"o}ckmann, Baumgardt,
\& Kroupa}{2010}]{lockmann:2010} L{\"o}ckmann U., Baumgardt H., Kroupa P., 2010, MNRAS, 402, 519

\bibitem[\protect\citeauthoryear{Maness et al.}{2007}]{maness:2007}
Maness H., et al., 2007, ApJ, 669, 1024

\bibitem[\protect\citeauthoryear{Mathis}{1967}]{mathis:1967} Mathis
J.~S., 1967, ApJ, 147, 1050

\bibitem[\protect\citeauthoryear{Melia}{1992}]{melia:1992} Melia
F., 1992, ApJ, 387, L25

\bibitem[\protect\citeauthoryear{Mo, van den Bosch,
\& White}{2010}]{mo:2010} Mo H., van den Bosch F.~C., White S., 2010, gfe..book,


\bibitem[\protect\citeauthoryear{Najarro et
al.}{1997}]{najarro:1997} Najarro F., Krabbe A., Genzel R., Lutz D., Kudritzki R.~P., Hillier D.~J., 1997, A\&A, 325, 700

\bibitem[\protect\citeauthoryear{Paumard et
al.}{2006}]{paumard:2006} Paumard T., et al., 2006, ApJ, 643, 1011

\bibitem[\protect\citeauthoryear{Quataert}{2004}]{quataert:2004}
Quataert E., 2004, ApJ, 613, 322

\bibitem[\protect\citeauthoryear{Rauch}{1999}]{rauch:1999} Rauch
K.~P., 1999, ApJ, 514, 725

\bibitem[\protect\citeauthoryear{Rybicki
\& Lightman}{1979}]{rybicki:1979} Rybicki G.~B., Lightman A.~P., 1979, rpa..book,


\bibitem[\protect\citeauthoryear{Salpeter}{1955}]{salpeter:1955}
Salpeter E.~E., 1955, ApJ, 121, 161

\bibitem[\protect\citeauthoryear{Sch{\"o}del et
al.}{2003}]{schodel:2003} Sch{\"o}del R., Ott T., Genzel R., Eckart
A., Mouawad N., Alexander T., 2003, ApJ, 596, 1015

\bibitem[\protect\citeauthoryear{Sch{\"o}del et
al.}{2002}]{schodel:2002} Sch{\"o}del R., et al., 2002, Nature, 419,
694

\bibitem[\protect\citeauthoryear{Sch{\"o}del et
al.}{2007}]{schodel:2007} Sch{\"o}del R., et al., 2007, A\&A, 469, 125

\bibitem[\protect\citeauthoryear{Spitzer}{1958}]{spitzer:1958}
Spitzer L., Jr., 1958, ApJ, 127, 17

\bibitem[\protect\citeauthoryear{Spitzer
\& Saslaw}{1966}]{spitzer:1966} Spitzer L., Jr., Saslaw W.~C., 1966, ApJ, 143, 400

\bibitem[\protect\citeauthoryear{Stolte et al.}{2005}]{stolte:2005}
Stolte A., Brandner W., Grebel E.~K., Lenzen R., Lagrange A.-M., 2005, ApJ,
628, L113

\end{thebibliography}
\end{document}